\documentclass[aps,twocolumn]{revtex4}
\makeatletter

\newcommand{\Rmnum}[1]{\expandafter\@slowromancap\romannumeral #1@}
\makeatother
\usepackage{graphicx}
\usepackage{dcolumn}
\usepackage{bm}
\usepackage{CJK}
\usepackage{color}
\usepackage{amsfonts}
\usepackage{psfrag}
\usepackage{wrapfig}
\usepackage{subfigure}
\usepackage{makeidx}
\usepackage{multirow}
\usepackage{epsf}
\usepackage[colorlinks,linkcolor=red,anchorcolor=red,citecolor=blue,urlcolor=blue]{hyperref}
\usepackage{amsmath}
\usepackage{cases}
\usepackage{bm}
\usepackage[dvipsnames]{xcolor}
\usepackage{lmodern}

\begin{document}
\title{Nondegenerate Akhmediev breathers and abnormal frequency jumping in multicomponent nonlinear Schr\"odinger equations}
\author{Shao-Chun Chen$^{1}$}
\author{Chong Liu$^{1,2,3}$}\email{chongliu@nwu.edu.cn}
\address{$^1$School of Physics, Northwest University, Xi'an 710127, China}
\address{$^2$Shaanxi Key Laboratory for Theoretical Physics Frontiers, Xi'an 710127, China}
\address{$^3$Peng Huanwu Center for Fundamental Theory, Xi'an 710127, China}

\begin{abstract}
Nonlinear stage of higher-order modulation instability (MI) phenomena in the frame of multicomponent nonlinear Schr\"odinger equations (NLSEs) are studied analytically and numerically.
Our analysis shows that the $N$-component NLSEs can reduce to $N-m+1$ components, when $m(\leq N)$ wavenumbers of the plane wave are equal.
As an example, we study systematically the case of three-component NLSEs which cannot reduce to the one- or two-component NLSEs.
We demonstrate in both focusing and defocusing regimes, the excitation and existence diagram of a class of nondegenerate Akhmediev breathers formed by nonlinear superposition between several fundamental breathers with the same unstable frequency but corresponding to different eigenvalues.
The role of such excitation in higher-order MI is revealed by considering the nonlinear evolution starting with a pair of unstable frequency sidebands.
It is shown that the spectrum evolution expands over several higher harmonics and contains several spectral expansion-contraction cycles.
In particular, abnormal unstable frequency jumping over the stable gaps between the instability bands are observed in both defocusing and focusing regimes. We outline the initial excitation diagram of abnormal frequency jumping in the frequency-wavenumber plane.
We confirm the numerical results by exact solutions of multi-Akhmediev breathers of the multi-component NLSEs.
\end{abstract}

\maketitle
\section{Introduction}
Modulation instability (MI) is a fundamental nonlinear phenomenon in various physical systems \cite{Bespalov,BF,Dudley1,Dudley2,Everitt_BEC,VSRB-2024}.
Nonlinear stage of MI evolving from periodic modulation on plane wave in the scalar focusing nonlinear Schr\"odinger equation (NLSE) exhibits either the conventional growth-decay cycle \cite{AB,Exp2001,Exp2009,Exp2018-1,Exp2018-2,TMP1987,MC1,MC2}
or more complex splitting wave dynamics \cite{JETP88,OC2010,Exp2011-OL,Exp2011-PRL,Exp2017-PRE}. The latter case is presently known as `higher-order MI' \cite{JETP88} containing multiple unstable modes, where each mode is associated with a corresponding `Akhmediev breather' (AB) \cite{AB}.
Specifically, nonlinear evolution of higher-order MI exhibits multi-ABs with continuous frequency jumping
provided that initial modulation frequency $\omega$ is below a critical low frequency limit and higher harmonics of modulation are also located
within the instability band.
Namely, the frequency jumping of multi-ABs follows
\begin{eqnarray}
\omega \rightarrow 2\omega\rightarrow 3\omega \rightarrow ... k\omega, ~~~~~~~~k\in \mathbb{N}.
\end{eqnarray}
Importantly, this complex dynamics has been confirmed experimentally in both optics \cite{Exp2011-PRL} and hydrodynamics \cite{Exp2017-PRE}.

In most complex situations, multi-component (vector) wave field rather than a single (scalar) one needs to
be considered \cite{OF,BEC,F}. Interaction between multiple components often induces new MI bands, which indicates nontrivial nonlinear stage of MI in these complex cases.
The mathematical model describing the multi-component wave field is typically a set of coupled nonlinear equations.
One of the most studied integrable models is the two-component NLSEs known as Manakov equations \cite{Manakov1974}.
Nonlinear coupling of the two components does induce new MI bands in both focusing and defocusing cases
\cite{VMI1987,VMI1991,OCW1,VMI2014} which eventually leads to new vector ABs with complex structures.
For example, vector MI can even exist in the defocusing regime when the two-component NLSEs admit different background wavenumbers \cite{VMI2014}.
The corresponding vector ABs can exhibit unique hidden dynamics in the nonlinear stage \cite{VH_DMI}.
One important finding is that such MI leads to the formation of rogue waves in the defocusing case \cite{VMI2014} which is absent in the scalar NLSE. This theoretical prediction has been observed experimentally in fiber optics \cite{RW_Vob1,RW_Vob2}.

On the other hand, vector MI of the focusing Manakov equations is no less surprising.
It is shown that the additional `X-shaped' MI band admits vector ABs with extremely asymmetric spectra \cite{VAB2021,VAB2022}.
Moreover, recent studies demonstrated a new class of nondegenerate ABs that consists of two fundamental ABs with the same unstable frequency but corresponding to different eigenvalues \cite{VAB2022}. It is found that such ABs can be excited from simple initial conditions in a wide range of physical parameters \cite{VAB2022}. Experimental observations of the spectral evolution of these vector ABs have recently been done in fiber optics \cite{L-exp2024}.
In addition, recent theoretical work shows that nonlinear evolution of higher-order MI starting from these initial conditions exhibits
abnormal frequency jumping over the stable gaps between the instability bands \cite{VH_FMI}.
Specifically, the frequency jumping of multi-ABs follows:
\begin{eqnarray}
\omega\rightarrow k\omega, ~~~~~~~~~~k\geq3;\\
\omega\rightarrow2\omega\rightarrow k\omega, ~~~~~~k\geq4.
\end{eqnarray}
Such abnormal frequency jumping occur only in the focusing Manakov system.
These findings reveal the richness of vector MI which is absent in the scalar NLSE.

The findings of vector MI mentioned above are limited to the two-component NLSEs.
Clearly, increasing the number of components of the model and exploring the variety of more complex MI evolution
are necessary and relevant.
In particular, recent experiment has realized properties of vector solitons in Bose-Einstein condensates based on the three-component NLSEs \cite{three-2020}. It is expected that theoretical results obtained in this paper could provide a basic in experimental observation of more complex MI excitations.
Yet, the complexity increases greatly when the number of components in the model exceeds two. Consequently, the corresponding nonlinear stage of vector MI remains unexplored largely. Indeed, such extension is far from simple even for vector solitons. These include the fundamental soliton \cite{G-2016}, soliton complexes \cite{Kanna_NLSE3,SC0,SC1,SC2} and nondegenerate solitons \cite{Qin_NLSE3}.
Recent results also demonstrated the characteristics of rogue waves \cite{VRW1,VRW2}, and Kuznetsov-Ma solitons \cite{VKM_DF} in the three-component NLSEs.
Although vector ABs produce highly nontrivial structures, the higher-order MI described by nonlinear superposition of multi-ABs in $N$-component NLSEs ($N>2$) has not be studied so far.

The paper is organized as follows.
We provide exact AB solutions of the $N$-component NLSEs and present the basic eigenvalue analysis of the solutions in Sec. \ref{Sec2}.
In Sec. \ref{Sec3}, we present a degeneracy analysis for the $N$-component NLSEs and further explore the MI variation laws in both the focusing and defocusing regimes.
Section \ref{Sec4} presents a class of nondegenerate ABs and its existence diagram for the three-component NLSEs.
Section \ref{Sec4-1} predicts unique higher-order MI dynamics involving abnormal frequency jumping.
In Section \ref{Sec5}, we perform direct numerical simulations that start from a single pair of sidebands to confirm the nonlinear evolution of  higher-order MI with abnormal frequency jumping of multi-ABs.
A particular case of the abnormal frequency jumping that exist only in the focusing three-component NLSEs is considered in details in Sec. \ref{Sec6}.
Section \ref{Sec7} shows the excitation diagram of such abnormal frequency jumping.
Section \ref{Sec8} contains our conclusions.

\section{$N$-component NLSEs and exact fundamental AB solutions}\label{Sec2}

The $N$-component nonlinear Schr\"odinger  equations \cite{Manakov1974}, in dimensionless form, are given by
\begin{equation}\label{eq1}
i\frac{\partial\psi^{(j)}}{\partial t}+\frac{1}{2}\frac{\partial^2\psi^{(j)}}{\partial x^2}+\sigma\left(\sum^N_{n=1}|\psi^{(n)}|^2\right)\psi^{(j)}=0,
\end{equation}
where $\psi^{(j)}(t,x)$ with $j=1,2,... N$ are the $N$ nonlinearly coupled components of the vector wave field. The physical meaning of independent variables $x$ and $t$ depends on a particular physical problem of interest.
We have normalised Eq. (\ref{eq1}) in a way such that $\sigma=\pm1$.
Note that in the case $\sigma=1$, Eqs. (\ref{eq1}) describe the focusing
(or anomalous dispersion) regime, in the case $\sigma=-1$,
Eqs. (\ref{eq1}) describe the defocusing (or normal dispersion) regime. When $N=2$, Eqs. (\ref{eq1}) denote the so-called Manakov equations.

Equations (\ref{eq1}) admit fundamental AB solutions in a unified form by using a Darboux transformation scheme \cite{Ling-2016}. They are given by \cite{VAB2021,VAB2022}
\begin{equation}
\psi^{(j)}=\psi_{0}^{(j)}\left[\frac{\cosh(\bm{\Gamma}+i\gamma_j)e^{i\eta_{1j}}+\varpi \cos(\bm{\Omega}-i\epsilon_j)e^{i\eta_{2j}}}{\cosh\bm{\Gamma}+\varpi \cos\bm{\Omega}}\right].\label{eqb}
\end{equation}
Here, $\psi_{0}^{(j)}$ denote the vector plane wave solution,
\begin{equation}
\psi_{0}^{(j)}=a_j \exp \left\{ i \left[{\beta_j}x + (\sum^N_{n=1}a_{n}^2- \beta_j^2/2) t \right] \right\}.\label{eqpw}
\end{equation}
Parameters $a_j$ and $\beta_j$ in (\ref{eqpw}) are the amplitude and the wavenumber of the $N$ plane wave components respectively.
The scalar arguments $\bm{\Gamma}$ and $\bm{\Omega}$ in (\ref{eqb}) are:
\begin{eqnarray}
\bm{\Gamma}=\omega \chi_i\bm t,~
\bm{\Omega}=\omega \left[\bm x+ (\chi_r+\frac{1}{2}\omega )\bm t \right]+\arg\frac{2\chi_i}{2\chi_i-i\omega}.
\end{eqnarray}
Here $\bm{x}=x-x_{1}$, $\bm{t}=t-t_{1}$ are shifted spatial and time variables respectively with  $x_{1}$ and $t_{1}$ being responsible for the spatial and temporal position of the centre of the breather.
Other notations in (\ref{eqb}) are:
\begin{eqnarray}
\eta_{1j}&=&\frac{\gamma_{1j}+\gamma_{2j}}{2},~~\eta_{2j}=\arg\frac{\chi^*+\beta_j}{\chi+\beta_j+\omega}, \\
\gamma_{j}&=&\frac{\gamma_{1j}-\gamma_{2j}}{2},~~\varpi= \Big|\frac{2\chi_i}{2\chi_i+i\omega} \Big|,\\
\gamma_{1j}&=&\arg\frac{\chi^*+\beta_j}{\chi+\beta_j},~~
\gamma_{2j}=\arg\frac{\chi^*+\beta_j+\omega}{\chi+\beta_j+\omega},\\
\epsilon_j&=&\log\left(\frac{(\chi^*+\beta_j)(\chi+\beta_j)}{(\chi+\beta_j+\omega)(\chi^*+\beta_j+\omega)}\right)^{1/2}.
\end{eqnarray}

An important parameter of the breather is its complex eigenvalue $\chi\equiv\chi(\sigma, a, \beta, \omega)$ with its real $\chi_r$ and imaginary $\chi_i$ parts. The constraint condition for $\chi$ are given by:
\begin{eqnarray}
1+\sum_{j=1}^N\frac{\sigma a_j^2}{(\chi+\beta_j)(\chi+\omega+\beta_j)}=0.\label{eqchi}
\end{eqnarray}
Once $\sigma$ is fixed, the AB solution (\ref{eqb}) depends on the background amplitudes $a_j$, the wavenumbers $\beta_j$, and the modulation frequency $\omega$.
It represents the full growth-decay cycle of MI. Namely, it
grows out of the plane wave (\ref{eqpw}) that is weakly modulated with frequency $\omega$.

Equation (\ref{eqchi}) is a $2N$-th order equation of eigenvalue $\chi$, where complex solutions are always paired complex conjugates.
Only imaginary part $\chi_i$ determines the gain rate of MI $G=|\omega \chi_i|$. Thus, purely real eigenvalues are invalid.
Solving (\ref{eqchi}) with $a_j$, $\beta_j$ for any $N$ to have explicit expressions of $\chi$ is not always possible. We present the eigenvalue expressions for some simple cases in Appendix \ref{chi-expression}. This helps to present our analysis below.
On the other hand, both the imaginary $\chi_i$ and real $\chi_r$ parts jointly determine the space-time structure of the fundamental ABs. This can be seen clearly by using the Hessian matrix analysis \cite{Ling-2016,VAB2021}. We present the details of the analysis in Appendix \ref{Hessian}.

For every pair of complex conjugated eigenvalues ($\chi$ and $\chi^*$), the AB solution (\ref{eqb}) satisfies a simple transformation
\begin{eqnarray}
\psi^{(j)}(\bm x, \chi)=\psi^{(j)}(\bm x+\Delta x,\chi^*)e^{i\Delta \phi_j},\label{eqpin}
\end{eqnarray}
where $\Delta x=\frac{2}{\omega}\left(\arg\frac{2\chi_i}{2\chi_i-i\omega}\right)$ is the shift along the $x$-axis and $\Delta \phi_j=2\eta_{1j}$ is the phase shift of the complex function, respectively. This means that $\psi^{(j)}(\chi)$ and $\psi^{(j)}(\chi^*)$ have the same amplitude profiles.
This is essentially the same solution but shifted along the $x$-axis.
For simplicity, we denote the solution satisfying Eq. (\ref{eqpin}) as $\psi^{(j)}(\chi)\Leftrightarrow\psi^{(j)}(\chi^*)$.

\section{Degeneracy analysis for the model}\label{Sec3}

Although we have the relation (\ref{eqpin}) of the AB solutions, analyzing the properties of ABs of the $N$-component NLSEs is still a tricky task when $N>2$. To better analyze the solutions, let's first consider the degeneracy of the $N$-component NLSEs.

For the AB solutions (\ref{eqb}), the wavenumbers $\beta_j$ play a key role in breather formation.
In particular, the choice of $\beta_j$ will lead to degeneracy of the model.
Specifically, two cases should be considered.

i) If all wavenumbers of the vector plane wave are equal, namely $\beta_1=\beta_2=...=\beta_N$, we have
\begin{eqnarray}
\psi^{(1)}/\psi^{(2)}/.../\psi^{(N)}=a_1/a_2/.../a_N.
\end{eqnarray}
Eq. (\ref{eq1}) degenerates into the scalar ($N=1$) NLSE as follows:
\begin{eqnarray}
i\frac{\partial\tilde{\psi}}{\partial t}+\frac{1}{2}\frac{\partial^2\tilde{\psi}}{\partial x^2}+\sigma|\tilde{\psi}|^2\tilde{\psi}=0,\label{eqsca}
\end{eqnarray}
where
\begin{eqnarray}
\tilde{\psi}=\frac{\sqrt{\sum^N_{j=1} a_n^2}}{a_1}\psi^{(1)}.\label{eqlar}
\end{eqnarray}
Reduction of the $N$-component NLSEs in this way results in only the results of the scalar NLSE.
Namely, MI is absent in the defocusing regime and abnormal frequency jumping of AB is absent in the focusing regime.

ii) If several ($m$) wavenumbers of the plane wave are equal ($m<N$), Eq. (\ref{eq1}) degenerates into a coupled NLSEs of $N-m+1$ components.
Namely, if we set $\beta_1=\beta_2=...=\beta_m$,
Eq. (\ref{eq1}) reduces to $(N-m+1)$-component NLSEs as follows:
\begin{eqnarray}
&&i\frac{\partial\tilde{\psi}}{\partial t}+\frac{1}{2}\frac{\partial^2\tilde{\psi}}{\partial x^2}+\sigma\left(|\tilde{\psi}|^2+\sum^N_{n=m+1}|\psi^{(n)}|^2\right)\tilde{\psi}=0,\nonumber\\
&&i\frac{\partial\psi^{(j)}}{\partial t}+\frac{1}{2}\frac{\partial^2\psi^{(j)}}{\partial x^2}+\sigma\left(|\tilde{\psi}|^2+\sum^N_{n=m+1}|\psi^{(n)}|^2\right)\psi^{(j)}=0,\nonumber\\ \label{eqzf}
\end{eqnarray}
where $m< j\leq N$, and
\begin{eqnarray}
\tilde{\psi}=\frac{\sqrt{\sum^m_{n=1} a_n^2}}{a_1}\psi^{(1)}.
\end{eqnarray}
Below, we apply Eqs. (\ref{eqzf})  to several special cases of the $N$-component NLSEs in both focusing and defocusing regimes.

\subsection{The defocusing regime}\label{Sec3.1}

For the defocusing regime ($\sigma=-1$) of the $N$-component NLSEs, Eq. (\ref{eqchi}) always yields a pair of real eigenvalues.
In particular, for the scalar defcosuing NLSE ($N=1$), there is only a pair of real eigenvalues. This means that MI is absent in the scalar defocusing NLSE system, $G=|\omega \chi_i|=0$. Only for the cases $N\geq2$, the multi-component interactions can induce MI in the defocusing regime
when the relative wavenumber exists.

\begin{figure}[!htb]
\centering
\includegraphics[width=85mm]{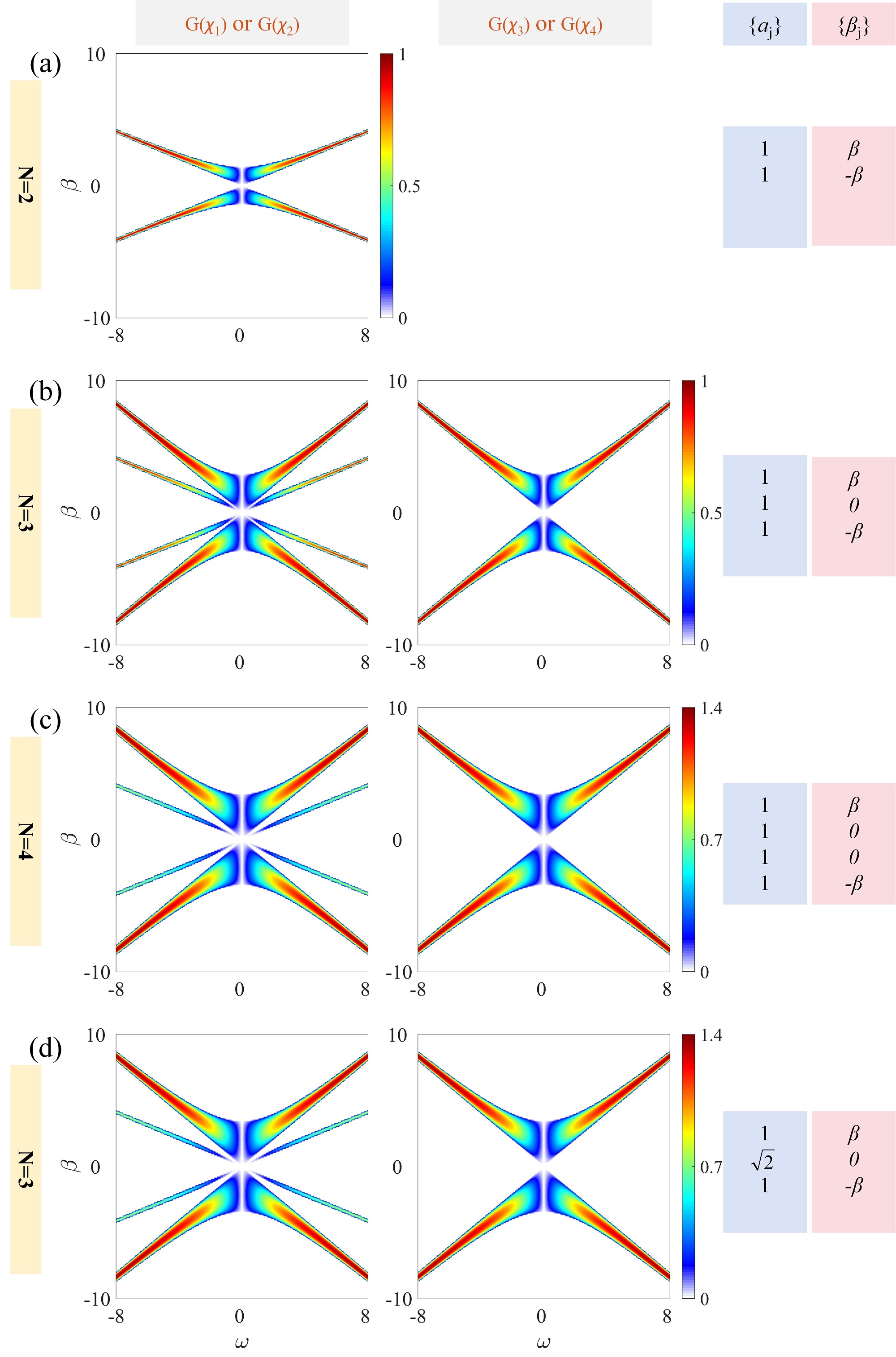}
\caption{MI growth rate $G=|{\omega} \chi_i|$ on the ($\omega,\beta$) plane given by the AB solution (\ref{eqb}) in  defocusing regime.
Parameter are shown in the figure.
} \label{f1}
\end{figure}

Figure \ref{f1} shows the characteristics of the defocusing MI growth rate $G=|\omega \chi_i|$ in ($\omega,\beta$) regime for cases of $N=2,3,4$.
Explicit expressions of all eigenvalues are given in Appendix \ref{chi-expression}.
The specific parameters of $a_j$ and $\beta_j$ are also shown in Fig. \ref{f1}.
For the case of $N=2$ with $a_j=1$ and $\{\beta_j\}=\{\beta,-\beta\}$, MI growth rate $G$ exhibits a X-shaped structure [see Fig. \ref{f1} (a)].
It is only given by $G(\chi_1)=G(\chi_2)$ (where $\chi_1=\chi_2^*$) since $\chi_3$ and $\chi_4$ are real parameters ($\chi_3,\chi_4\in\mathbb{R}$).
When $N$ increases, MI distributions become more complex.
As shown in Fig. \ref{f1} (b), for the case $N=3$ with $a_j=1$ and $\{\beta_j\}=\{\beta,0,-\beta\}$, we have multiple MI growth rates rather a single one for the case $N=2$. Namely, we have two growth rates given by $G(\chi_1)=G(\chi_2)$ and $G(\chi_3)=G(\chi_4)$.
In particular, for the case $G(\chi_1)$, one more pair of MI branches are observed.
Such new MI bands lead to nontrivial frequency jumps in nonlinear
stage of vector higher-order modulation instability, see Sec. \ref{Sec5}.

\begin{figure*}[!htb]
\centering
\includegraphics[width=130mm]{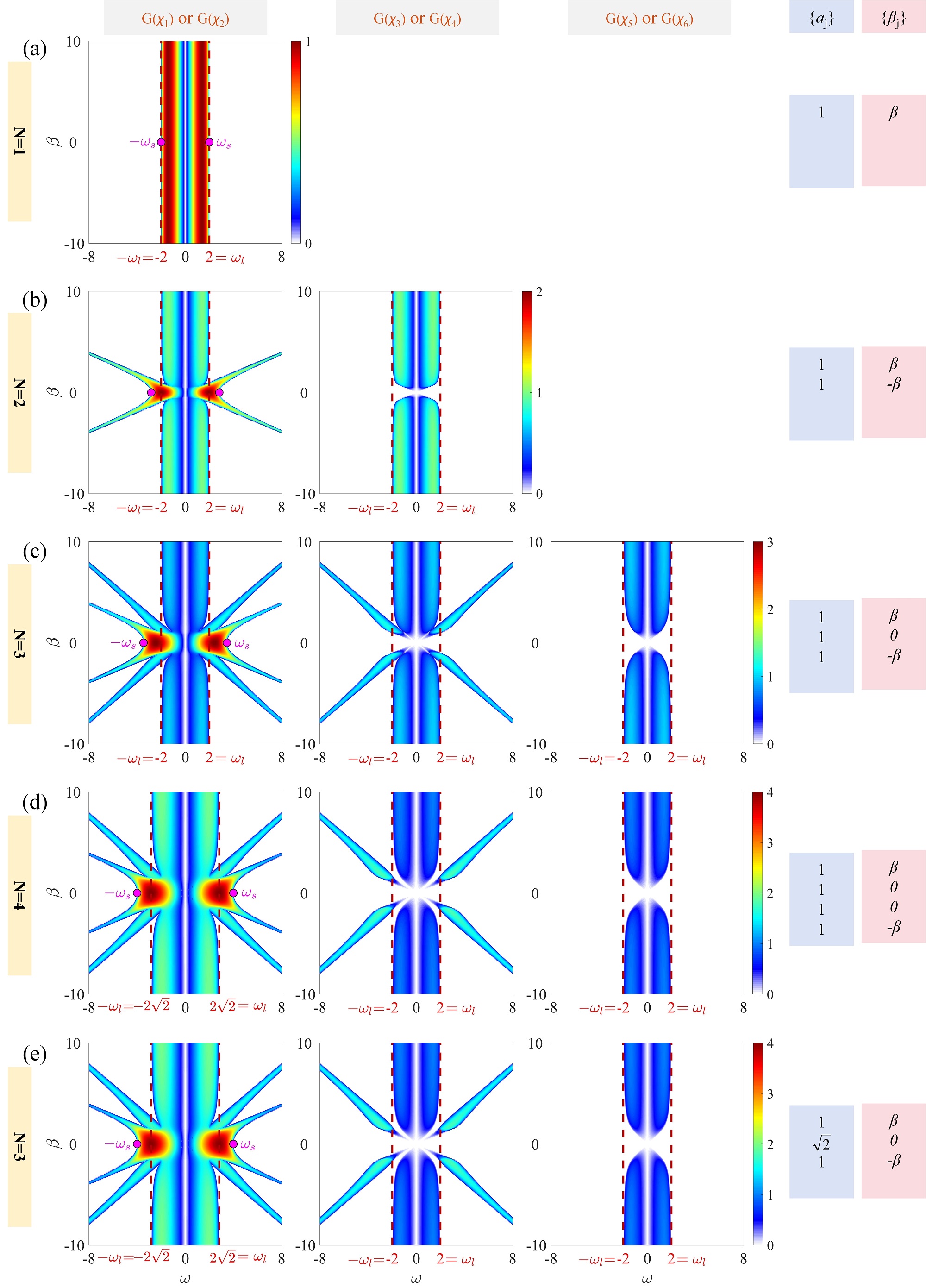}
\caption{MI growth rate $G=|{\omega} \chi_i|$ on the ($\omega,\beta$) plane given by the AB solution (\ref{eqb}) in  focusing regime.
Parameter are shown in the figure.
} \label{f2}
\end{figure*}

\begin{figure*}[!htb]
\centering
\includegraphics[width=130mm]{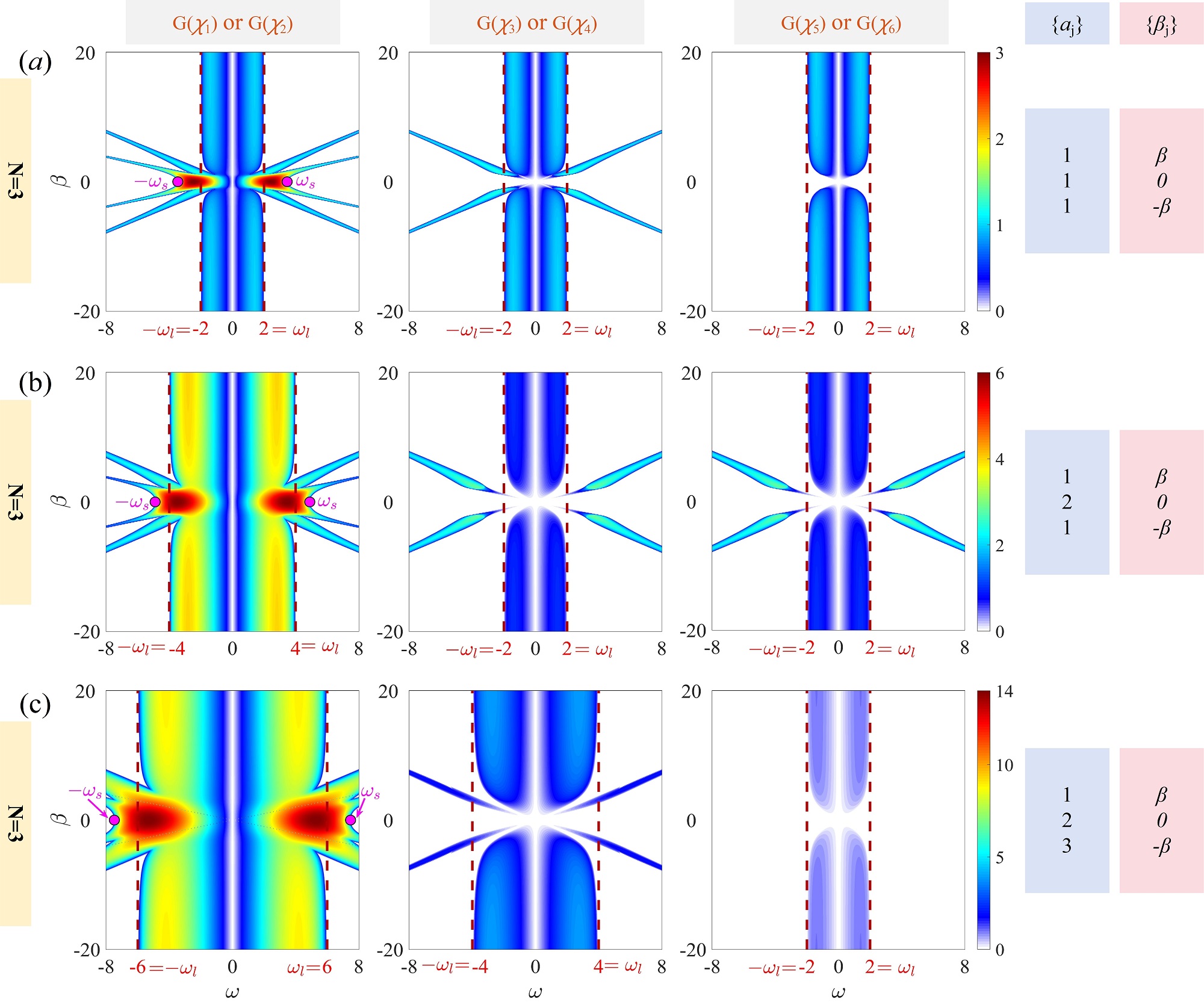}
\caption{MI growth rate $G=|{\omega} \chi_i|$ on the ($\omega,\beta$) plane in the focusing ($\sigma=1$) three-component system  with $\{\beta_j\}=\{\beta,0,-\beta\}$ and unequal background amplitude $a_1\neq a_2\neq a_3$. (a) $\{a_j\}=\{1,1,1\}$; (b) $\{a_j\}=\{1,2,1\}$; (c) $\{a_j\}=\{1,2,3\}$.
} \label{f3}
\end{figure*}

For the case $N=4$, complexity of MI patterns increases when all wavenumbers are different. However, if we consider two of them are equal, i.e.,  $\{\beta_j\}=\{\beta,0,0,-\beta\}$, the corresponding MI patterns shown in Fig. \ref{f1}(c) are similar to those of $N=3$ shown in Fig. \ref{f1}(b).
In fact, from (\ref{eqzf}) we know that the 4-component NLSEs with $\beta_2=\beta_3$ (implying $\psi^{(2)}/\psi^{(3)}=a_2/a_3$) reduce to the 3-component NLSEs of form:
\begin{eqnarray}
i\frac{\partial\tilde{\psi}}{\partial t}+\frac{1}{2}\frac{\partial^2\tilde{\psi}}{\partial x^2}+\sigma(|\tilde{\psi}|^2+|\psi^{(1)}|^2+|\psi^{(4)}|^2)\tilde{\psi}=0,\nonumber\\
i\frac{\partial\psi^{(j)}}{\partial t}+\frac{1}{2}\frac{\partial^2\psi^{(j)}}{\partial x^2}+\sigma(|\tilde{\psi}|^2+|\psi^{(1)}|^2+|\psi^{(4)}|^2)\psi^{(j)}=0,\nonumber\\
\label{eq342}
\end{eqnarray}
where $j=1,4$, and
\begin{eqnarray}
&&\tilde{\psi}=\frac{\sqrt{a_2^2+a_3^2}}{a_2}\psi^{(2)}.\label{eq341}
\end{eqnarray}
From (\ref{eq342}) we know that the MI growth rates show in Fig. \ref{f1} (c) with $\{\beta_j\}=\{\beta,0,0,-\beta\}$ and $a_j=1$ turns out to be these of 3-component NLSEs (\ref{eq342}) with $\{\beta_j\}=\{\beta,0,-\beta\}$ and $a_j=\{1,\sqrt{2},1\}$.
Fig. \ref{f1} (d) illustrates the MI growth rate $G$ for the 3-component NLSEs (\ref{eq342}).
Clearly, the MI distributions are the same as the case of 4-component NLSEs shown in Fig. \ref{f1} (c).

\subsection{The focusing regime}\label{Sec3.2}

Unlike the defocusing cases, the focusing MI distribution ($\sigma=1$) is more complex.
For the scalar focusing NLSE ($N=1$), the MI is limited to the frequency range $(-2a_1, 2a_1)$ [see Fig. \ref{f2}(a)].
For $N\geq2$, we have two MI variation laws depending on $\beta_j$.

(1) When $\beta_1=\beta_2=...=\beta_N$, Eq. (\ref{eq1}) is equivalent to a scalar NLSE system.
In this case, the range of the MI band is $(-\omega_s, \omega_s)$, where
\begin{eqnarray}
\omega_s=2\sqrt{a_1^2+a_2^2+...+a_N^2} \label{eqws}.
\end{eqnarray}
When $\omega=\sqrt{2(a_1^2+a_2^2+...+a_N^2)}$, we have the maximum value of the growth rate
\begin{eqnarray}
G_{max}=a_1^2+a_2^2+...+a_N^2 \label{eqGmax}.
\end{eqnarray}

(2) When $\beta_1\neq\beta_2\neq...\neq\beta_N$, there exists a main MI band that does not vary with relative wavenumber $\beta$. It can be seen from the red dashed lines in Fig. \ref{f2}. The main MI band is limited to the range of $(-\omega_{l},\omega_{l})$..
According to Eq. (\ref{eqchi}), we can obtain the explicit expression of $\omega_{l}$ as $\beta\rightarrow\pm\infty$ as follows:
\begin{eqnarray}
\{\omega_{l,1},...\omega_{l,N}\}=2\times \mathrm{one}~\mathrm{of}~\{a_1, a_2,..., a_N\} .\label{eqwl1}
\end{eqnarray}
Namely, $\pm\omega_{l}$ of each type of main MI bands associated with a pair of complex conjugate eigenvalues is twice one of the background amplitudes $\{a_j\}$.
For the scalar focusing NLSE ($N=1$), we have $\omega_s=\omega_l=2a_1$.
As shown in Fig. \ref{f2} (a), the purple dots fall on the red dashed lines.

Figure \ref{f2} illustrates the characteristics of the focusing MI growth rate $G=|\omega \chi_i|$ in ($\omega,\beta$) regime for different $N$ ($N=1,2,3,4$). The specific parameters of $a_j$ and $\beta_j$ are also shown in Fig. \ref{f2}.
In all plots, when $\beta_j=0$, the MI band is limited to $(-\omega_s, \omega_s)$, see the purple dots in the leftmost panels of Fig. \ref{f2};
when $\beta\rightarrow\pm\infty$, the main MI bands is limited to $(-\omega_l, \omega_l)$, given by Eq. (\ref{eqwl1}).

Just like the defocusing case, as $N$ increases, more MI bands appear in the focusing case.
For $N=2$, two MI growth rates $G(\chi_1)=G(\chi_2)$ and $G(\chi_3)=G(\chi_4)$ are obtained.
Only one of them, say $G(\chi_1)$, exhibits an additional X-shaped MI band.
For $N=3$, three MI growth rates can be seen from Fig. \ref{f2}(c). Growth rate $G(\chi_1)$ has two additional X-shaped MI bands.
Note that only $G(\chi_1)$ can reduce to the scalar NLSE growth rate when $\beta=0$.
Otherwise, the growth rate is vanishing when $\beta=0$.
Furthermore,
the MI growth rates shown in Fig. \ref{f2} (c) with $\{\beta_j\}=\{\beta,0,0,-\beta\}$ and $a_j=1$ turns out to be these of 3-component NLSEs (\ref{eq342}) with $\{\beta_j\}=\{\beta,0,-\beta\}$ and $a_j=\{1,\sqrt{2},1\}$.
This is once again confirmed by degeneracy analysis of Eq. (\ref{eq342}).

To further verify the validity of Eqs. (\ref{eqws}) and (\ref{eqwl1}), we display in Fig. \ref{f3} MI distributions of 3-component focusing NLSE system with different background amplitudes $a_j$.
As seen from Fig. \ref{f3}, both the $\omega_s$ (purple points) and the maximum values of the color bars confirm the Eqs. (\ref{eqws}) and (\ref{eqGmax}), respectively.
To further illustrate Eq. (\ref{eqwl1}), here we take three components with different background amplitudes as an example. In the three-component system, Eq. (\ref{eqwl1}) can be further expressed as
\label{eqwl}\begin{eqnarray}
\omega_{l}(\chi_1, \chi_2)&=&2~\max\{a_1, a_2, a_3\},\nonumber\\
\omega_{l}(\chi_3, \chi_4)&=&2~\textrm {med}\{a_1, a_2, a_3\},\label{eqwl3}\\
\omega_{l}(\chi_5, \chi_6)&=&2~\min\{a_1, a_2, a_3\}.\nonumber
\end{eqnarray}
Here, med is the function that takes the middle value, i.e., $\min\{a_1, a_2, a_3\}\leq\textrm {med}\{a_1, a_2, a_3\}\leq\max\{a_1, a_2, a_3\}$.
Fig. \ref{f3} shows three cases of three-component MI: (a) all three amplitudes are equal, (b) two amplitudes are equal, and (c) all three amplitudes are different.
The MI main band frequency boundaries $\omega_l$ all satisfy Eq. (\ref{eqwl3}) (or Eq. (\ref{eqwl1})). They are shown by red dashed lines in Fig. \ref{f3}.

\section{Nondegenerate ABs and their existence diagram in three-component NLSEs}\label{Sec4}

\begin{figure}[htb]
\centering
\includegraphics[width=86mm]{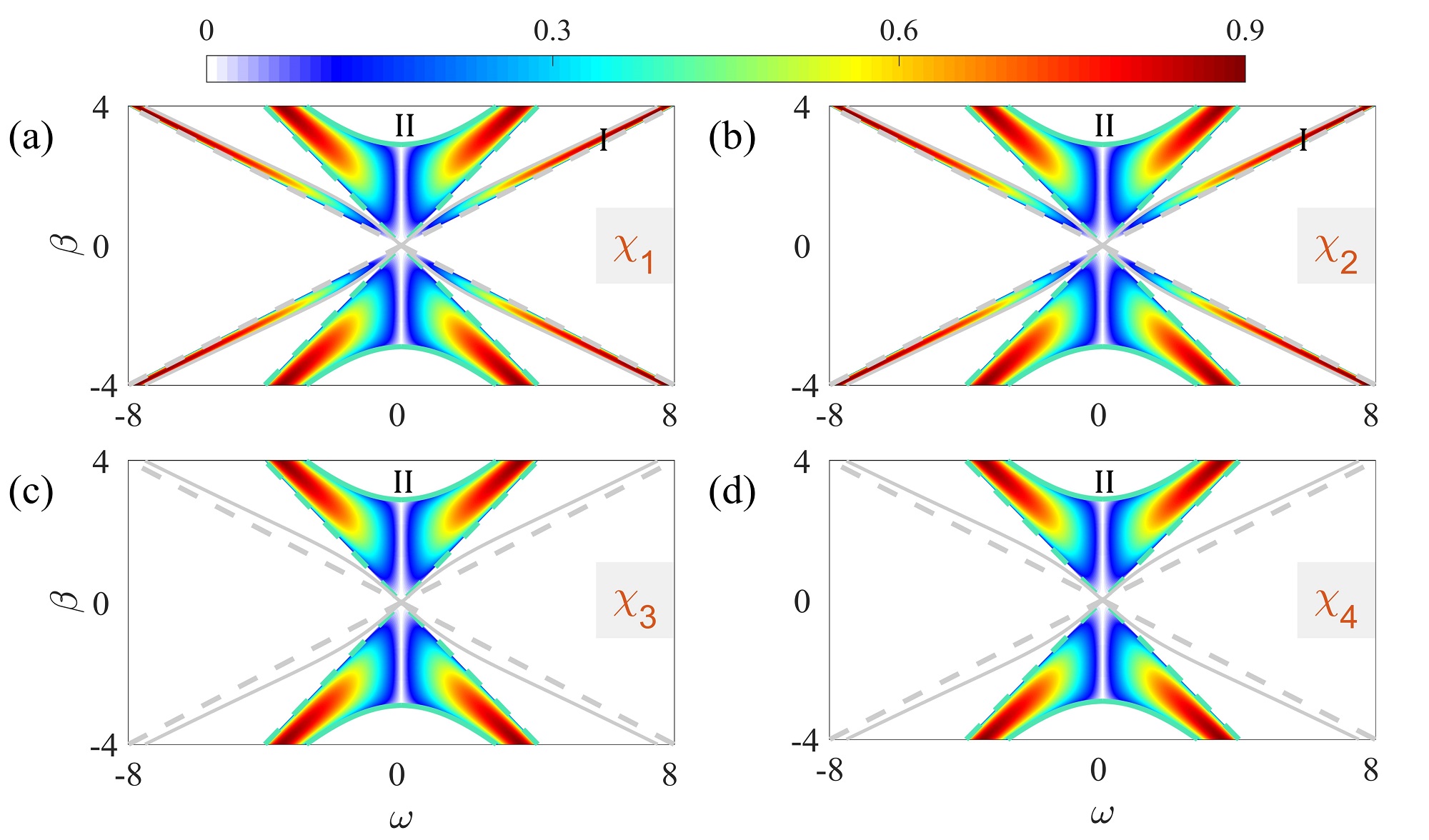}
\caption{MI growth rate $G=|{\omega} \chi_i|$ on the ($\omega,\beta$) plane given by the AB solution (\ref{eqb}) in  defocusing regime.
Parameter $a=1$.}\label{f4}
\end{figure}

\begin{figure}[htb!]
\centering
\includegraphics[width=88mm]{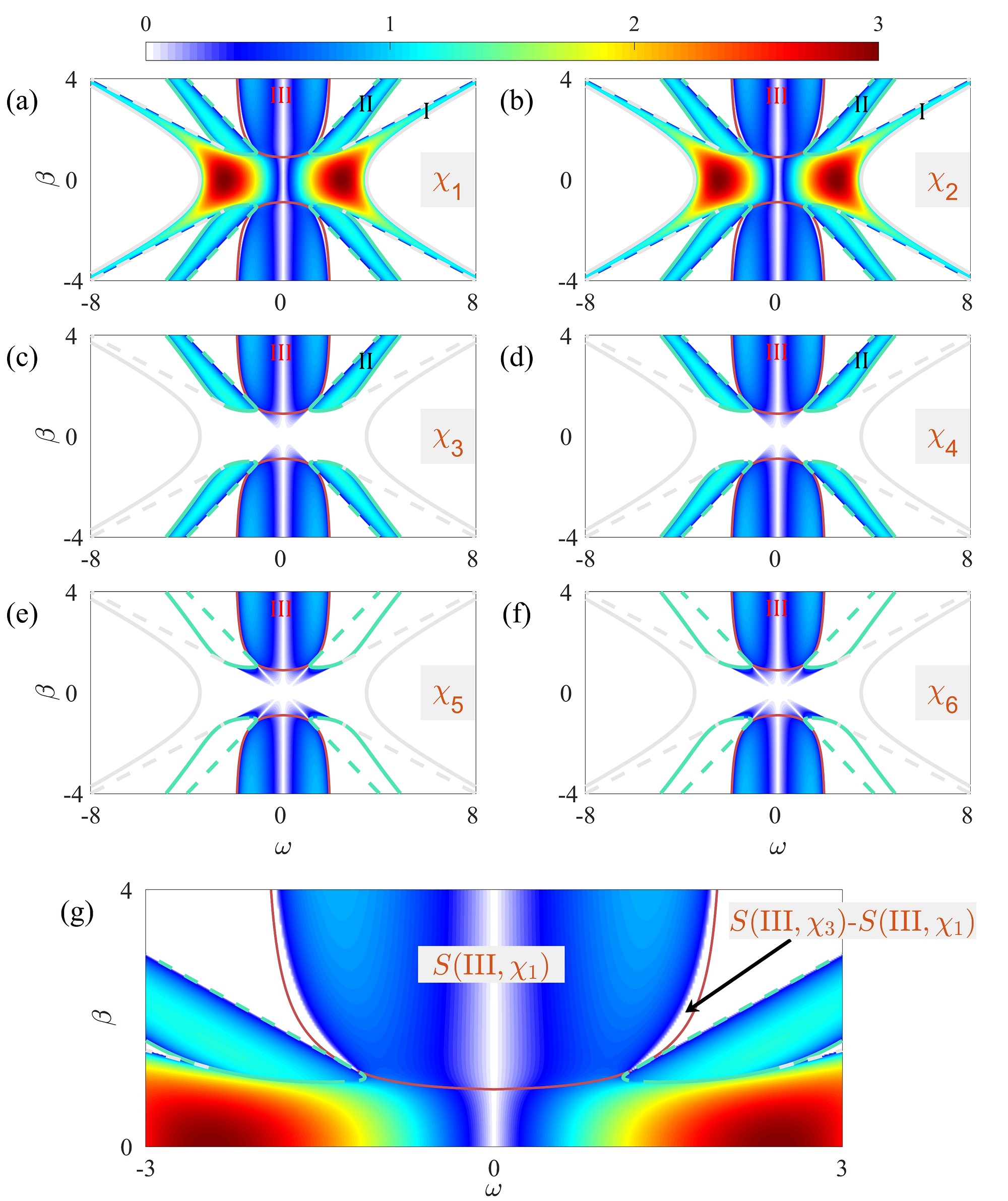}
\caption{MI growth rate $G=|{\omega} \chi_i|$ on the ($\omega,\beta$) plane given by the AB solution (\ref{eqb}) in focusing regimes.
(g) the magnified version of (a), where region (III) is divided into two distinct cases: \emph{case A} $S_A=S(\rm{III},\chi_1)$; \emph{case B} $S_B=S(\rm{III},\chi_3)-S(\rm{III},\chi_1)$.
Parameter $a=1$.
}\label{f5}
\end{figure}

Based on the above analysis, study of MI of $N$-component NLSE ($N>2$) is of great significance and necessity,
as it can provide a more comprehensive and in-depth understanding of the MI dynamics.
Next, we focus our attentions on the case of 3-component NLSEs and consider the characteristics of AB solutions.
Without loss of generality, we first consider a symmetric framework associated with the parameters of the plane wave background:
\begin{equation}\label{fra1}
\begin{split}
a_1=a_2=a_3=a,~~ \\
\beta_1=-\beta_3=\beta,~\beta_2=0.
\end{split}
\end{equation}
Equation (\ref{fra1}) leads to the symmetry of the vector wave field $\psi^{(3)}(\beta)=\psi^{(1)}(-\beta)$.
The corresponding MI growth rates are shown in Figs. \ref{f4} and \ref{f5}. The AB solutions that exist on the same diagram should be analyzed carefully.
Moreover, by substituting Eq. (\ref{fra1}) into Eq. (\ref{eqchi}), we can obtain the eigenvalues. The specific form is presented in Appendix \ref{chi-expression}.

\subsection{The defocusing regime}\label{Sec4.1}

Let us analyze first the AB solutions on the plane of MI growth rate $G=|{\omega} \chi_i|$  in the defocusing case shown in Fig. \ref{f4}.
When $\sigma=-1$, Eq. (\ref{eqchi}) always has a pair of real roots, i.e., $\chi_{5i}=\chi_{6i}=0$.
Moreover, $G(\chi_1)=G(\chi_2)$ and $G(\chi_3)=G(\chi_4)$.
We identify two (I and II) regions in the figures.
Based on the condition (\ref{eqchi}) and the relation (\ref{eqpin}), we conclude that:\\
(1) In the region (I), we have $\chi_1^*=\chi_2$, which indicates that the AB solutions satisfy $\psi^{(j)}(\chi_1)\Leftrightarrow\psi^{(j)}(\chi_2)$, see (\ref{eqpin}).
Thus, only one AB $\psi^{(j)}(\chi_1)$ [or $\psi^{(j)}(\chi_2)$] solution exists in this area.\\
(2) In the region (II), we have $\chi_1^*=\chi_3$, $\chi_2^*=\chi_4$, and $\chi_{1r}\neq\chi_{2r}$.
Thus, $\psi^{(j)}(\chi_1)\Leftrightarrow\psi^{(j)}(\chi_{3})$ and $\psi^{(j)}(\chi_2)\Leftrightarrow\psi^{(j)}(\chi_{4})$.
There are two ABs $\{\psi^{(j)}(\chi_1),\psi^{(j)}(\chi_{2})\}$ [or $\{\psi^{(j)}(\chi_3)$, $\psi^{(j)}(\chi_{4})\}$] in region (II), which can be verified by the analysis of AB structures (via the Hessian matrix analysis) shown in Fig. \ref{f13}.
Nonlinear superpositions between these two solutions result in the so-called nondegenerate second-order AB solutions \cite{VAB2022}.
Two-component NLSEs have no analogue in the defocusing regime. We will show their role in nonlinear stage of modulation instability in the 3-component NLSEs, see Sec. \ref{Sec5}.

\subsection{The focusing regime}\label{Sec4.2}

We then consider the AB solutions in the focusing case $\sigma=1$.
Three specific regions can be identified from the distribution of the MI growth rate shown in Fig. \ref{f5}:
(I) the X-shaped region limited by the grey solid  curves and four dashed straight lines,
(II) the regions limited by the four green curves and four green dashed straight lines,
(III) two U-shaped regions limited by the two red curves.

(1) In the region (I), there is one AB solution $\psi^{(j)}(\chi_1)$ [or $\psi^{(j)}(\chi_2)$].
Namely, $\psi^{(j)}(\chi_1)\Leftrightarrow\psi^{(j)}(\chi_2)$ since $\chi_1^*=\chi_2$.
Other cases ($\chi_3\sim\chi_6$) are invalid since they cannot reduce to the scalar NLSE AB when $\beta=0$.

(2) In the region (II), we have $\chi_1^*=\chi_3$, $\chi_2^*=\chi_4$, and $\chi_{1r}\neq\chi_{2r}$.
Thus, $\psi^{(j)}(\chi_1)\Leftrightarrow\psi^{(j)}(\chi_{3})$ and $\psi^{(j)}(\chi_2)\Leftrightarrow\psi^{(j)}(\chi_{4})$.
There are two ABs $\{\psi^{(j)}(\chi_1),\psi^{(j)}(\chi_{2})\}$ [or $\{\psi^{(j)}(\chi_3)$, $\psi^{(j)}(\chi_{4})\}$] in region (II). This is consistent with Hessian matrix analysis shown in Figs. \ref{f14} (a)-(d).

(3) In the region (III),  we identify two subregions which admit different numbers of AB solutions. They are subregion A: $S_A=S(\rm{III},\chi_1)$ and subregion B: $S_B=S(\rm{III},\chi_3)-S(\rm{III},\chi_1)$.
For subregion A, there are three AB solutions $\{\psi^{(j)}(\chi_1), \psi^{(j)}(\chi_3), \psi^{(j)}(\chi_4)\}$ [or $\{\psi^{(j)}(\chi_2), \psi^{(j)}(\chi_5), \psi^{(j)}(\chi_6)\}$].
For subregion B, there are two AB solutions $\{\psi^{(j)}(\chi_3), \psi^{(j)}(\chi_4)\}$ [or $\{\psi^{(j)}(\chi_5), \psi^{(j)}(\chi_6)\}$].

\begin{figure}[htb]
\centering
\includegraphics[width=88mm]{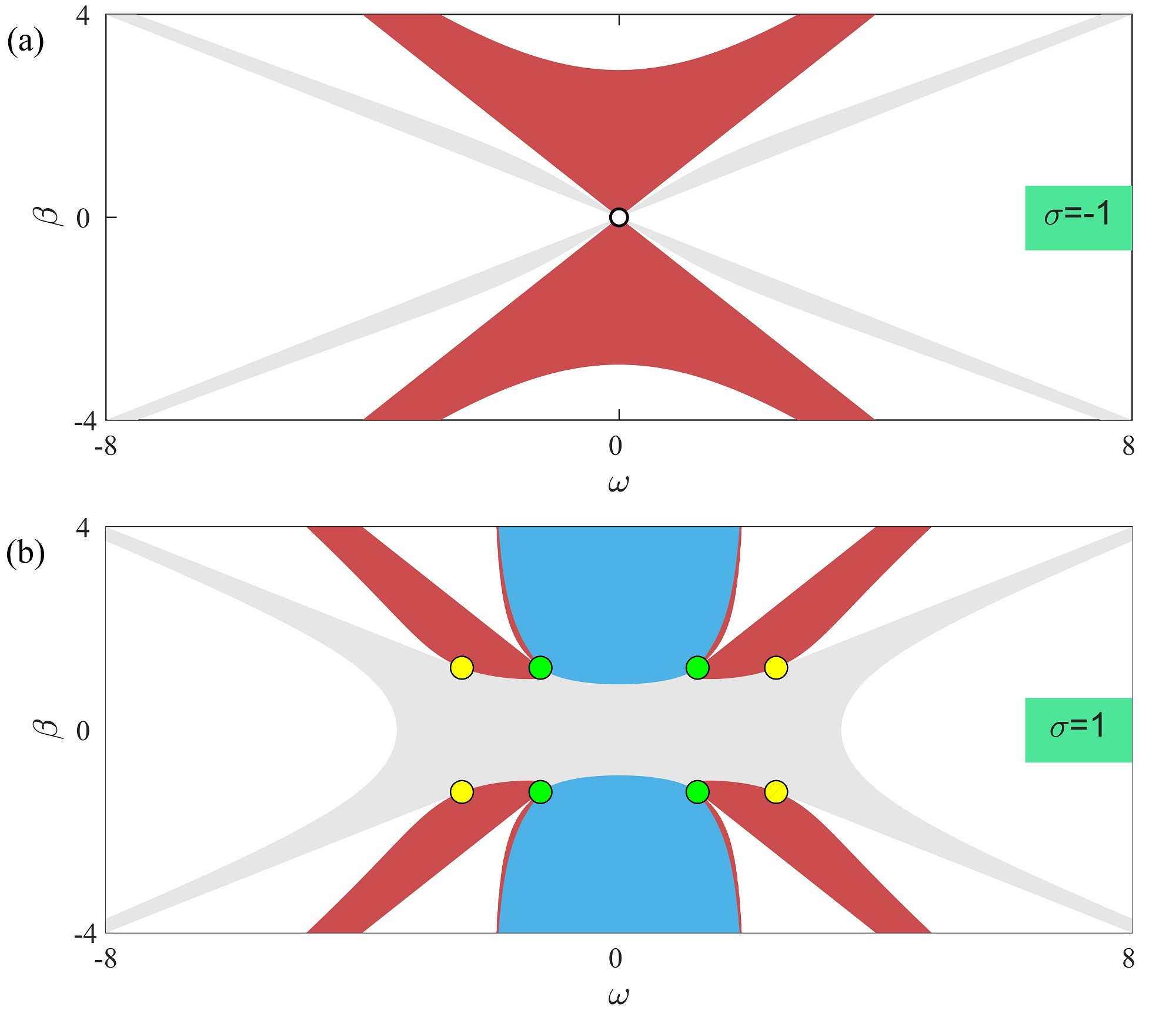}
\caption{The existence diagram of the number of ABs  on the ($\omega,\beta$) plane in (a) defocusing and (b) focusing regimes. The grey, red and blue areas correspond to one, two and three ABs, respectively.
The four large green and four large yellow dots in (b) are the branch points given by (\ref{eq-bp1}) and (\ref{eq-bp2}).
Parameter $a=1$.}\label{f6}
\end{figure}

Based on the analysis above, we obtain the existence diagram for the nondegenerate ABs on the ($\omega,\beta$) plane shown in Fig. \ref{f6}.
The grey, red, and blue areas correspond to one, two and three ABs, respectively.
The focusing case is more complex than the defocusing case.
One example is the branch points between different regions.
As can be seen from the figures, there is only one branch point (the white dot) for the defocusing case:
\begin{eqnarray}
(\omega_b,\beta_b)=(0,0).\label{eq-bp0}
\end{eqnarray}
While the branch points (green and yellow dots) of the focusing case are given by
\begin{eqnarray}
&&(\omega_b,\beta_b)=(\pm\sqrt{6}a/2,\pm\sqrt{6}a/2),\label{eq-bp1}\\
&&(2\omega_b,\beta_b)=(\pm\sqrt{6}a,\pm\sqrt{6}a/2).\label{eq-bp2}
\end{eqnarray}
Having these branch points we can predict nontrival nonlinear stage of higher-order MI when the initial parameters
$\beta\neq0$ (for the defocusing case) and $\beta^2>3a^2/2$ (for the focusing case).
Namely, abnormal frequency jumping over the stable gaps between the instability bands are observed
in both defocusing and focusing regimes.

\section{Prediction of higher-order MI dynamics from existence diagram}\label{Sec4-1}

\begin{figure}[!htb]
\centering
\includegraphics[width=86mm]{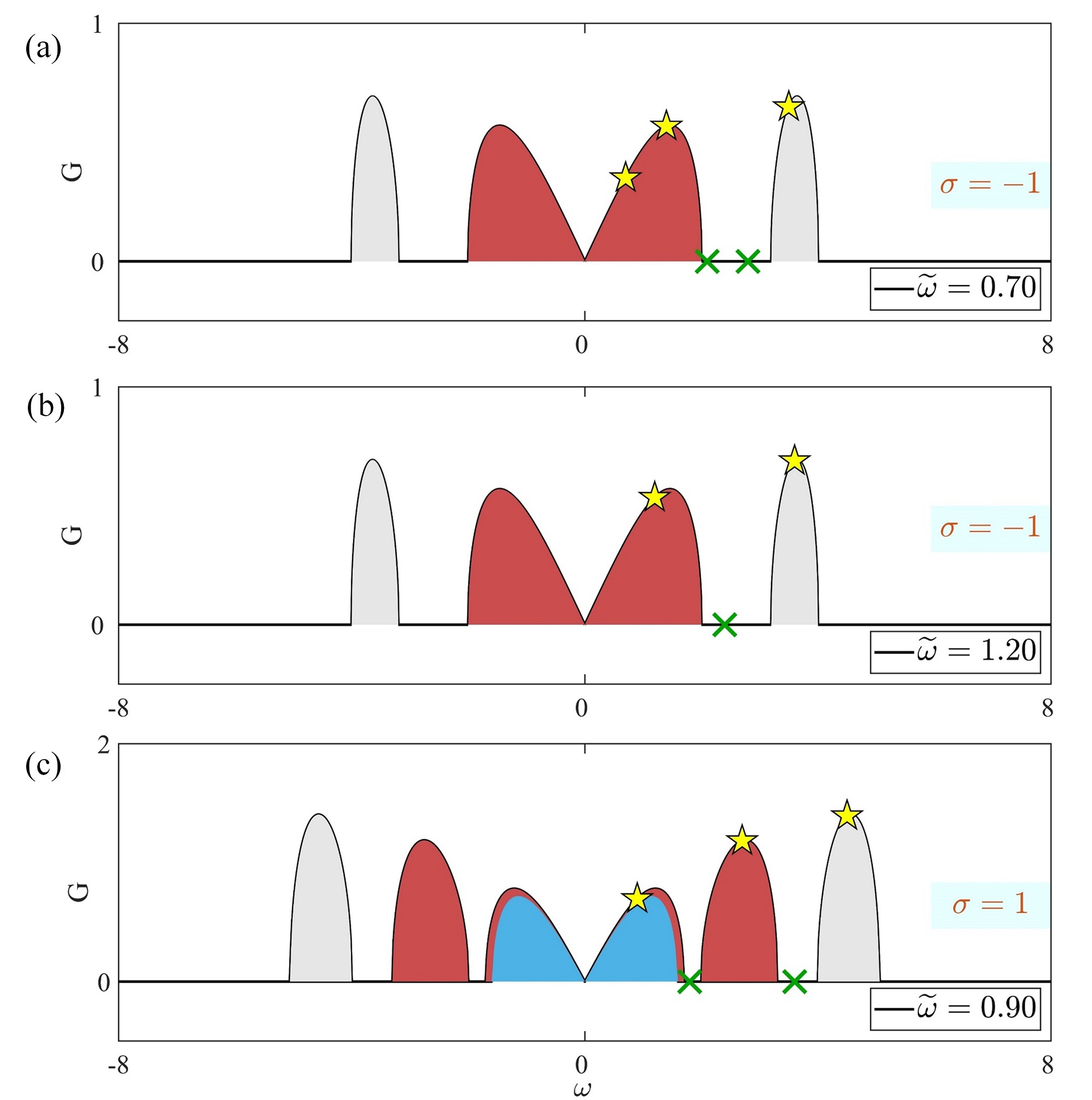}
\caption{Three examples of the MI growth rate spectra for $\beta=2$ in the (a)-(b) defocusing and (c) focusing cases.
Grey areas correspond to a single AB solution at each frequency. Red areas contain two different AB solutions at each frequency.
Blue areas contain three different AB solutions at each frequency. In all cases, $a=1$.}\label{f7}
\end{figure}

In general, higher-order MI can be excited from initial modulation involving any multiple unstable sidebands.
The nonlinear evolution thus involves multiple unstable modes (ABs).
However, the first sideband of the induced MI spectrum can be chosen arbitrarily in experiments.
The choice of the first sideband of the initial modulation influences the full-scale evolution that starts with the MI.
One simple way to excite higher-order MI is to use the simple modulation only involving the first sideband where the frequency $\omega$ is below a critical low frequency limit and higher harmonics of modulation are also located within the instability band.
Below, we will show how the choice of the first sideband can result in the nontrivial nonlinear evolution of higher-order MI.

\subsection{The defocusing regime}\label{}

In the defocusing regime, we consider the case $\beta\neq0$ ($\beta=2$). Then two examples with different frequencies of the first sideband are shown in Figs. \ref{f7}(a) and (b).

Figure \ref{f7}(a) shows the first sideband chosen to be $\tilde{\omega}=0.7$.
This frequency and its second harmonic $2\tilde{\omega}$ are located in the red lobe. They are shown by the yellow stars on the red lobe.
The latter is close to the maximum of the growth rate.
However, the two higher harmonics $3\tilde{\omega}$ and $4\tilde{\omega}$  fall into the spectral gap between the two lobes and remain stable. They are shown by the green crosses in Fig. \ref{f7}(a).
There are no ABs at these frequencies. On the contrary, the fifth harmonic $5\tilde{\omega}$ appears to the left of the maximum of of the grey lobe.
It is shown by the yellow star on the grey lobe. It is unstable and the corresponding single AB solution does exist. Four ABs can be excited within the red lobe of the spectrum and one AB can be excited in the grey area. Thus, the full scale evolution will involve five ABs.

In the case shown in Fig. \ref{f7}(c), the value of $\tilde{\omega}$ is even higher ($\tilde{\omega}=1.2$).
There are two different ABs that correspond to this frequency.
The second harmonic falls in the gap, shown by the green cross in Fig. \ref{f7}(c).
There is no growing AB at this frequency.
Moreover, the third harmonic $3\omega$ falls slightly close to the maximum of the grey lobe of the spectrum [yellow star in Fig. \ref{f7}(c)]. There is a single AB corresponding to this frequency.
Thus, the full scale MI evolution will involve three ABs.

\begin{figure*}[!htb]
\centering
\includegraphics[width=130mm]{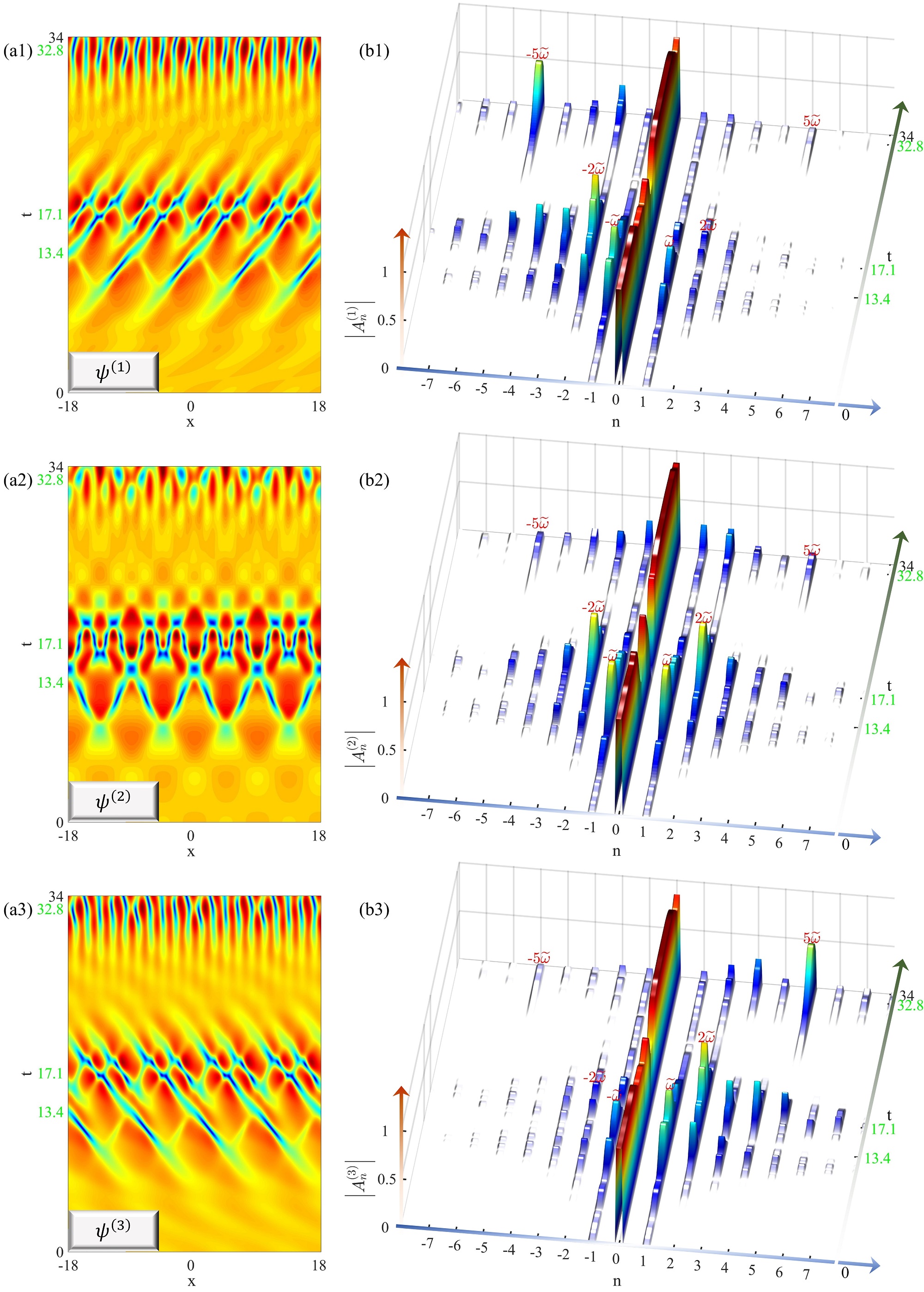}
\caption{Higher-order MI evolution with the growth rate spectrum shown in Fig. \ref{f7}(b).
Parameters $\tilde{\omega} = 0.7$, $\beta = 2.0$ and $\sigma=-1$.
(a) The results of numerical simulations started from initial conditions \eqref{eqin}.  (b) The evolution of the discrete spectrum for the same simulations.
} \label{f8}
\end{figure*}

\subsection{The focusing regime}\label{}

In the focusing case, we consider the case $\beta^2>\beta_b^2(=3a^2/2)$, where the MI growth rate spectrum splits into three lobes with two stable gap between them at each side. We fix $\beta=2$ and consider the frequency of the first sideband $\tilde{\omega}=0.9$ in Fig. \ref{f7}(c).
In this case, the first sideband and its two higher harmonics with frequencies $3\tilde{\omega}$, $5\tilde{\omega}$ are on the blue, red and gray lobes respectively. They are shown in Fig. \ref{f7}(d) by the yellow stars.
Each of them is close to the maximum of the growth rate of each lobe.
On the other hand, the second and fourth harmonics ($2\tilde{\omega}$, $4\tilde{\omega}$) fall into two different stable gaps, respectively.
They are shown by the green crosses in Fig.\ref{f7} (d).  There are no growing ABs at these frequencies.
Thus, the higher-order MI in this case will produces six ABs including three ABs at $\tilde{\omega}$, two ABs at $3\tilde{\omega}$ and one ABs at $5\tilde{\omega}$.
This can result in a new abnormal frequency jumping, which is absent in the three-component defocusing case and in the two-component focusing case.

\begin{figure*}[!htb]
\centering
\includegraphics[width=130mm]{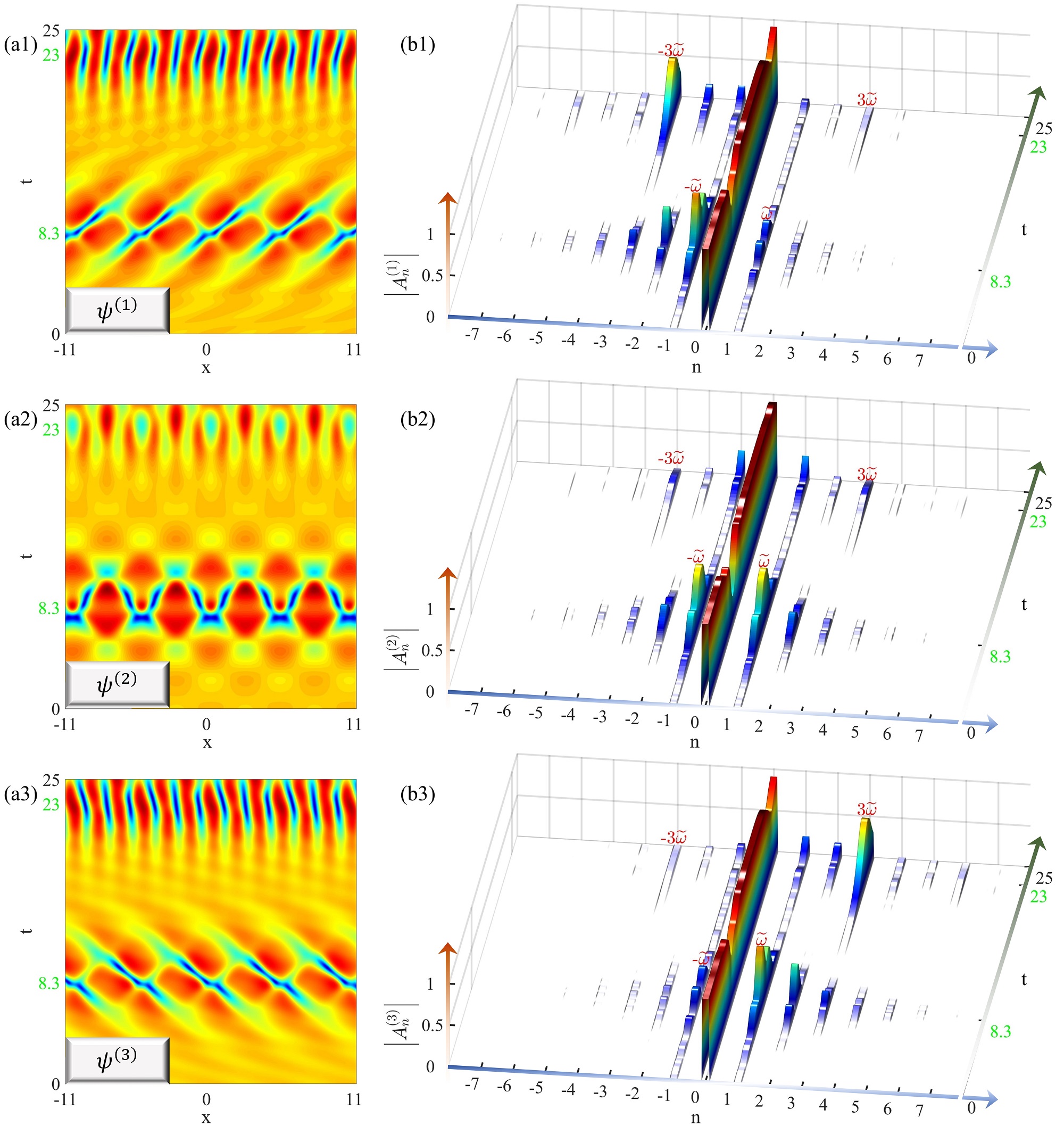}
\caption{Higher-order MI evolution with the growth rate spectrum shown in Fig. \ref{f7}(c).
Parameters $\tilde{\omega} = 1.2$, $\beta = 2.0$ and $\sigma=-1$.
(a) The results of numerical simulations started from initial conditions \eqref{eqin}. (b) The evolution of the discrete spectrum for the same simulations.
} \label{f9}
\end{figure*}

\section{Numerical simulations of higher-order vector MI dynamics}\label{Sec5}

\begin{figure*}[!htb]
\centering
\includegraphics[width=130mm]{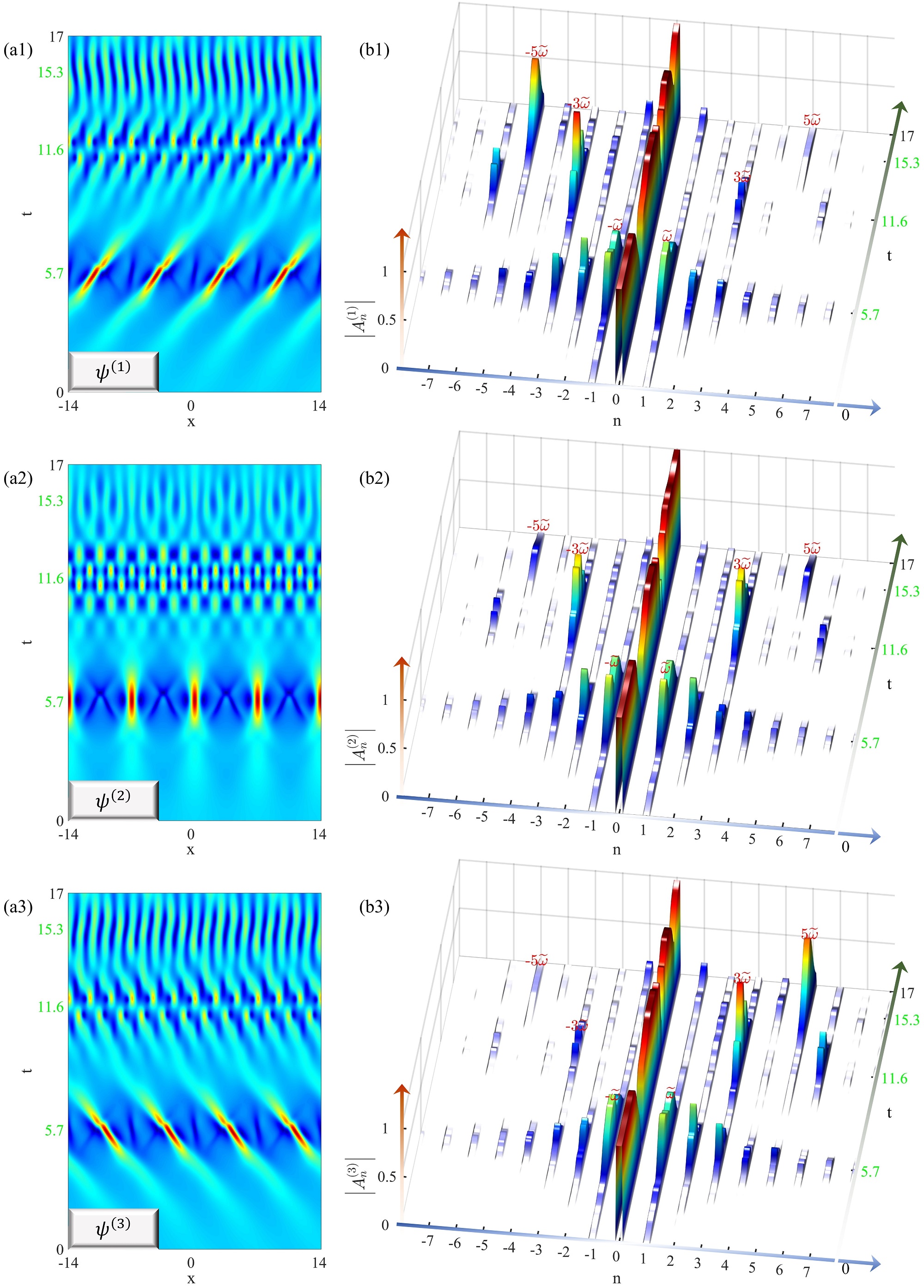}
\caption{Higher-order MI evolution with the growth rate spectrum shown in Fig. \ref{f7}(b).
Parameters $\tilde{\omega} = 0.9$, $\beta = 2.0$ and $\sigma=1$.
(a) The results of numerical simulations started from initial conditions \eqref{eqin}. (b) The evolution of the discrete spectrum for the same simulations.
} \label{f10}
\end{figure*}

Below, we perform simulations by integrating numerically the Manakov equations (\ref{eq1}) to confirm the above predictions.
The initial conditions we used do not correspond to the exact mathematical form of solutions.
Instead, we use a simple initial first sideband nodulation:
\begin{equation}
\psi^{(j)}=(1+\varepsilon\cos\tilde{\omega} x)\psi_{0}^{(j)},\label{eqin}
\end{equation}
where $\varepsilon$ ($\ll1$) denotes a small amplitude of modulation with a single frequency $\tilde{\omega}$, and $\psi_{0}^{(j)}$ is the plane wave background (\ref{eqb}).
The results of the numerical simulations corresponding to Figs. \ref{f7} (a)-(c) are shown in Figs. \ref{f8}-\ref{f10}.
Remarkably, all of these numerical results can be reproduced well by the exact higher-order AB solutions.
Among the numerical simulations shown in Figs. \ref{f8}-\ref{f10}, we only confirmed here the most complex higher-order MI dynamics shown in Figs. \ref{f10} by the exact sixth-order AB solutions [see Fig. \ref{f11}].

Figure \ref{f8} shows the higher-order MI dynamics when the spectrum of the MI growth rate and the modulation frequency are the same as in Fig. \ref{f7}(b). The results of numerical simulations for the evolution of the wave profile are shown in Fig. \ref{f8}(a). The corresponding evolution of the spectra is shown in Fig. \ref{f8}(b).
The components of the discrete spectrum are numbered by the integer $n$.
The number $n=0$ corresponds to the pump mode while $|n|\geq1$ numbers the sidebands [$n=\pm1, \pm2, ... \pm7$].
The selected values of $t$ are shown in green fonts in Figs. \ref{f8}(a). The choice of these values of $t$ is dictated by the points of maximal energy transfer from the pump to the sidebands.  These figures contain three growth-return cycles of the higher-order MI.

The positive and negative modes sidebands in the wave field $\psi^{(1)}$ and $\psi^{(3)}$ are asymmetric with respect to the pump mode, i.e., $|A_{n}^{(1)}|\neq|A_{-n}^{(1)}|$, $|A_{n}^{(3)}|\neq|A_{-n}^{(3)}|$.
Moreover, the corresponding $n$th mode of the AB in $\psi^{(1)}$ wave field is exactly opposite to that in the $\psi^{(3)}$ wave field.
Namely, $|A_n^{(3)}|=|A_{-n}^{(1)}|$, as can be seen from Figs. \ref{f8}(b1) and \ref{f8}(b3).
On the contrary, in the $\psi^{(2)}$ component, energy in the $n$th mode is equal to energy in the $(-n)$th mode. Namely, the spectra of the $\psi^{(2)}$ remain symmetric at any $t$, $|A_{n}^{(2)}|=|A_{-n}^{(2)}|$.

The initial modulation grows exponentially into breathers with period $2\pi/\tilde{\omega}$. The breathers then split into subwaves with period $\pi/\tilde{\omega}$. After that, these subwaves split into smallamplitude breathers with smaller period $2\pi/(5\tilde{\omega})$.
As the breathers of the first two cycles exist in the region (II), each cycle consists of two different ABs. Namely, the first cycle corresponds to the ABs with the frequency $\omega=\tilde{\omega}$; the second cycle consists of the ABs with the frequency $\omega=2\tilde{\omega}$.
This corresponds to the breather splitting with normal frequency jumping ($\tilde{\omega}\rightarrow2\tilde{\omega}$) in the region (II).
On the other hand, the breather of the third cycle corresponds to the region (I), which is defined by a single AB with the frequency $\omega=5\tilde{\omega}$ rather than the AB at frequency $3\tilde{\omega}$  or $4\tilde{\omega}$.
This is because the third and fourth harmonics are stable modes that do not produce AB.
This corresponds to the breather splitting with abnormal frequency jumping ($2\tilde{\omega}\rightarrow5\tilde{\omega}$). This can be confirmed by the spectrum evolution shown in Fig. \ref{f8}(b). As can be seen, only the $n=\pm5$ sideband is well enhanced at the third expansion-contraction cycle.

Figure \ref{f9} shows the higher-order MI dynamics when the second harmonic of the spectral components is located in stable gap of the growth rate spectrum as shown in Figs. \ref{f7}(b).
Numerical results for the wave evolution excited by the initial conditions (\ref{eqin}) with $\tilde{\omega}=1.2$ and $\beta=2$ are displayed in Fig. \ref{f9}(a).
The corresponding spectrum is shown in Fig. \ref{f9}(b).
The higher-order MI in this example exhibits only two growth-decay cycles.
The initial modulation in Fig. \ref{f9}(a) develops into the ABs with the transverse period $2\pi/\tilde{\omega}$ and the maximum modulation at $t=8.3$.
This structure further evolves into a breather with the period $2\pi/(3\tilde{\omega})$.
Each minimum of the previous breather splits into the three smaller minima rather than two.
Correspondingly, the spectrum evolution shown in Fig. \ref{f9}(b) reveals the enhancement of the third-order sidebands ($\pm3\tilde{\omega}$) at the second expansion-contraction cycle.
Instead, the second-order sidebands ($\pm2\tilde{\omega}$) are completely suppressed.
The first cycle consists of two different ABs with the same frequency $\tilde{\omega}$. The second cycle can be described by the AB with the frequency $3\tilde{\omega}$. This corresponds to the breather evolution with abnormal frequency jumping ($\tilde{\omega}\rightarrow3\tilde{\omega}$).

\section{A complex case of higher-order MI and its exact solutions}\label{Sec6}

\begin{figure*}[!htb]
\centering
\includegraphics[width=130mm]{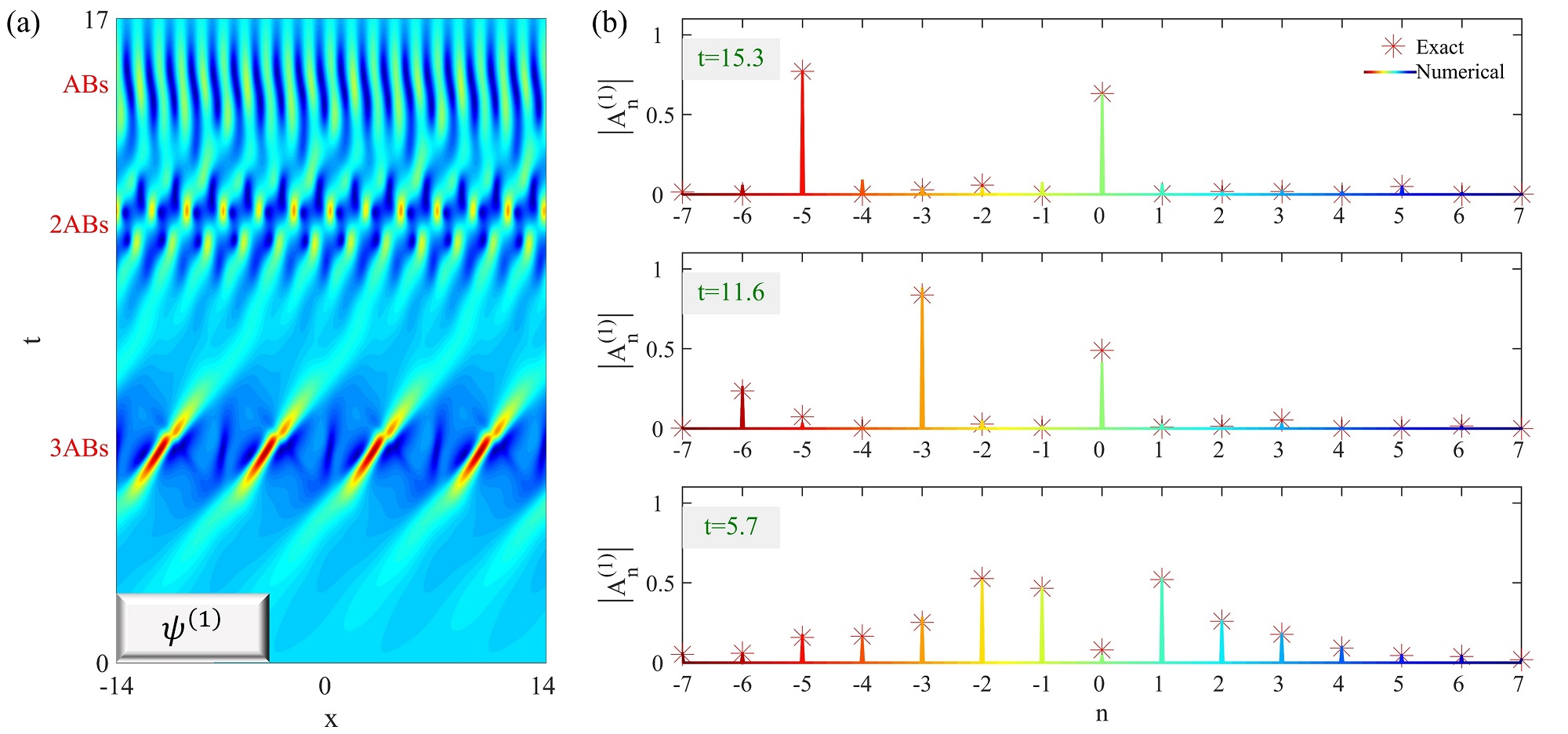}
\caption{
(a) The exact 6th-order AB solution confirming numerical simulations in Fig. \ref{f10}. Parameters of the solution are:
$\chi_{\textrm{set}}=\{-0.4500+0.7386 i, 1.4670+0.7737 i, -2.3670+0.7737 i, -2.3293+0.4382 i, -0.3707+0.4382 i, -2.2500+0.3094 i\}$,
$\omega_{\textrm{set}}=\{\tilde{\omega}, \tilde{\omega}, 2\tilde{\omega}, 2\tilde{\omega}, 6\tilde{\omega}\}$, $\Delta x_{\textrm{set}}=\{-3.12, -1.77, 1.41, -0.98, -3.48, -1.50\}$ and $\Delta t_{\textrm{set}}=\{7.05, 6.80, 6.10, 11.40, 12.20, 13.70\}$.
(b) Discrete spectra of the wave field obtained from the numerical simulations (vertical bars)  and from the exact solution (crosses) at three selected values of $t$ shown in green fonts in Figs.\ref{f10} (a)-(d).
The other parameters are the same as Fig. \ref{f10}.
} \label{f11}
\end{figure*}

The above results only involve a single stable gap between the instability bands. Let then consider a more complex case that involves two stable gaps just as shown in Fig. \ref{f7} (c).
Figure \ref{f10} shows the corresponding higher-order MI revolution in both time and frequency domains.
As can be seen, three growth-decay cycles are shown in Figs. \ref{f10}.
The initial modulation produces the AB with period $2\pi/\tilde{\omega}$ at $t=5.7$ at the expense of the pump.
The initial ABs split into the ABs with period $2\pi/(3\tilde{\omega})$ at $t=11.6$, and even further split into the AB with period $2\pi/(5\tilde{\omega})$ at $t=15.3$.
On the other hand, the $n=\pm2$ and $n=\pm4$ modes are suppressed in the next two cycles without producing the corresponding AB.

These three cycles exist in the region (III), the region (II) and the region (I) respectively, as shown in Fig. \ref{f7} (d).
Specifically, the first cycle consists of three different ABs with the same frequency $\tilde{\omega}$. The second cycle can be described by two ABs with the same frequency $3\tilde{\omega}$. This means that the first abnormal frequency jumping ($\tilde{\omega}\rightarrow3\tilde{\omega}$) of the breather evolution occurs in the first two cycles.
The third cycle is characterized by one AB with frequency $5\tilde{\omega}$.
Accordingly, the second abnormal frequency jumping ($3\tilde{\omega}\rightarrow5\tilde{\omega}$) exists in the last two cycles.
This can be confirmed by the spectra evolution shown in Fig. \ref{f10}(b).
As can be seen, the three instability modes $n=\pm1, \pm3, \pm5$ are well enhanced at three selected moments, respectively.

The numerical evolution of higher-order MI with abnormal frequency jumping ($\tilde{\omega}\rightarrow3\tilde{\omega}\rightarrow5\tilde{\omega}$) involve six fundamental ABs. Exact sixth-order AB solutions, in principle can describe the full nonlinear evolution of the numerical simulation. However,
the parameters in the exact solutions should be chosen carefully. Here, the sixth-order solution $\psi^{(j)}_{M=6}$ involves six unstable frequencies $\omega_{\textrm{set}}=\{\omega_1,...,\omega_M\}=\{\tilde{\omega}, \tilde{\omega}, \tilde{\omega}, 3\tilde{\omega}, 3\tilde{\omega}, 5\tilde{\omega}\}$. In addition, the spatiotemporal patterns of such multi-ABs depend on the relative separations in both $x$ and $t$, i.e., $\Delta x_{\textrm{set}}=\{x_{1}, ..., x_{M}\}$ and $\Delta t_{\textrm{set}}=\{t_{1}, ..., t_{M}\}$. These parameters are given in the caption of Fig. \ref{f11}.

Figure \ref{f11}(a) shows the amplitude distributions of the sixth-order solution (for simplicity only $\psi^{(1)}$ is shown here).
As can be seen, the exact solution is in good agreement with the numerical simulations in Fig. \ref{f10}(a1). The discrete spectra obtained from the numerical simulations (vertical bars) and the exact results (stars) at selected values of $t$ are shown in Fig. \ref{f11}(b).
The selected values of $t$ here correspond to the maximal energy transfer from the carrier wave to the sidebands.
Comparison of these spectra also shows good agreement between the numerical simulations and exact results.
Indeed, when parameters are correctly chosen, the exact AB solution reproduces well the numerical results in the two other cases discussed above.

\begin{figure}[htb]
\centering
\includegraphics[width=86mm]{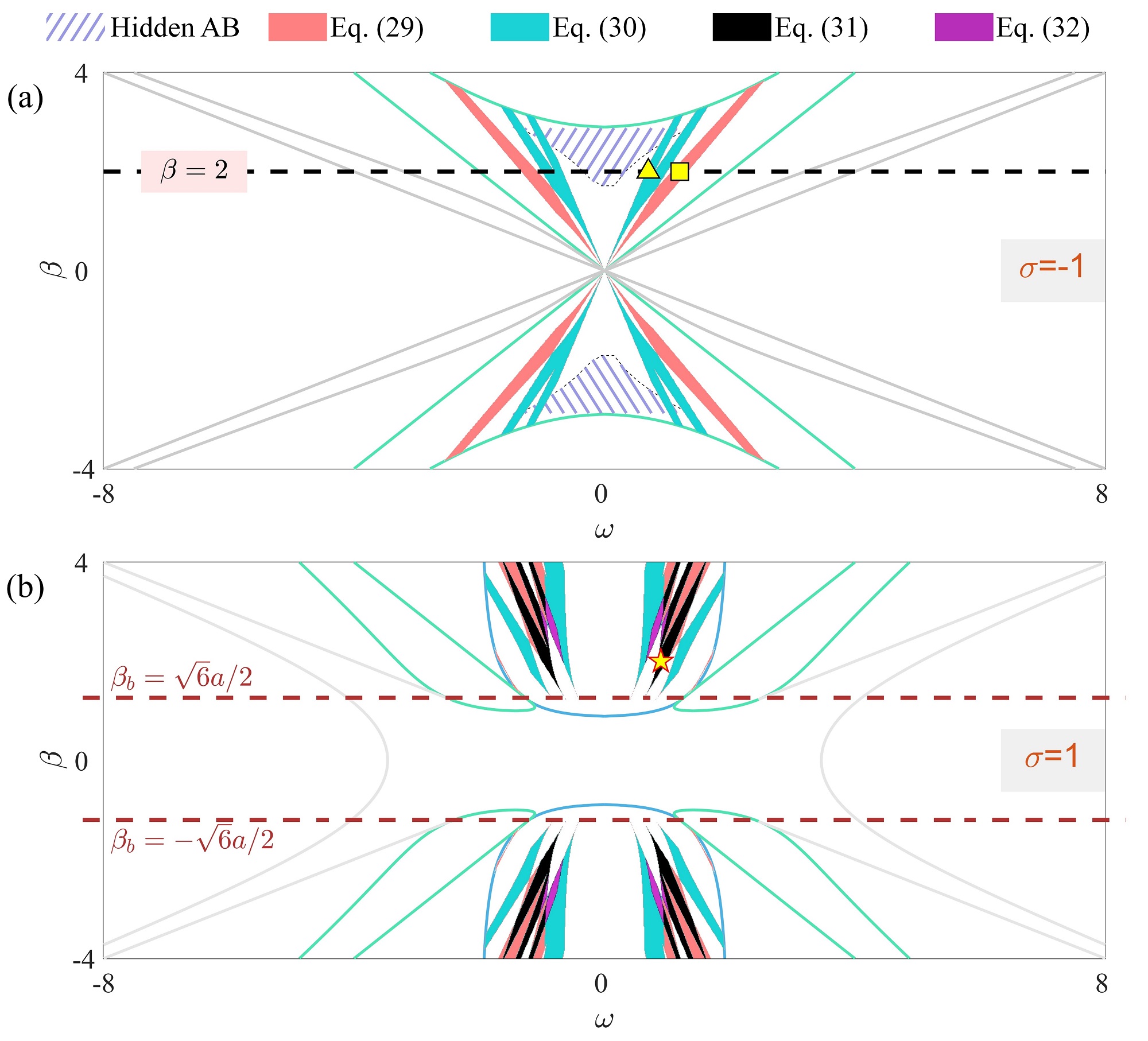}
\caption{The ($\omega,\beta$) plane of initial conditions.  The higher-order MI with two types of frequency jumping can be excited in the cyan and pink areas.
The yellow  triangle and square corresponds to the initial parameters that we used to generate higher-order MI dynamics shown in Figs. \ref{f8} and \ref{f9}, while the yellow star shows the parameters used for generation of results shown in Fig. \ref{f10}.}\label{f12}
\end{figure}

\section{Excitation diagram of abnormal frequency jumping}\label{Sec7}

Numerical simulations of higher-order MI dynamics shown in Figs. \ref{f8}-\ref{f10} are only particular examples involving multi-ABs with abnormal frequency jumping.
In fact, we have found that such MI dynamics can be excited from a wide range of the parameters of initial conditions \eqref{eqin} in both focusing and defocusing cases. In Fig. \ref{f7}, we present the existence diagram of such excitations on the ($\omega,\beta$) plane obtained numerically.
In either focusing or defocusing regime, such higher-order MI dynamics can be excited when the parameters $\omega$, and $\beta$ are in the cyan, pink, black and purple areas.

Let us first consider the defocusing case shown in Fig. \ref{f12} (a).
Existence diagram of two types of abnormal frequency jumping (pink and cyan areas) are found numerically in this case.
Higher-order MI in the pink area involves the energy transfer between the spectral components that are not the nearest neighbours, i.e.
\begin{eqnarray}
\tilde{\omega}\rightarrow k\tilde{\omega}, ~~~~~~~~~~k\geq3. \label{eq-J1}
\end{eqnarray}
A particular case with $k=3$ is shown in Fig. \ref{f9}.
The parameters of the initial conditions that correspond to this case are represented by the yellow square in Fig. \ref{f12}. We call this direct abnormal frequency jumping of multi-ABs.
On the other hand, higher-order MI in the cyan area involves, as the first step, energy transfer between the closest components ($\tilde{\omega}\rightarrow2\tilde{\omega}$). The second step is the energy transfer across the spectral components with  $n\geq4$.
Namely,
\begin{eqnarray}
\tilde{\omega}\rightarrow 2\tilde{\omega}\rightarrow k\tilde{\omega}, ~~~~~~~~~~k\geq4.\label{eq-J2}
\end{eqnarray}
The corresponding MI dynamics involving such spectral jumping when $k=5$ is shown in Fig. \ref{f8}. The parameters of the initial conditions that correspond to this case are represented by the yellow  triangle in Fig. \ref{f12}.
We call this indirect abnormal frequency jumping.
Note that the higher-order MI in the shaded area involves `hidden AB' phenomenon \cite{VH_DMI} during nonlinear evolution.
This is consistent with the two-component defocusing NLSE system \cite{VH_DMI}.

In the focusing regime, existence diagram of four types of abnormal frequency jumping (pink, cyan, purple, black areas) are found numerically.
This can be clearly seen from Fig. \ref{f12} (b).
Just like the defocusing case, the pink and cyan areas correspond to abnormal frequency jumping described by Eq. (\ref{eq-J1}) and (\ref{eq-J2}), respectively.
Higher-order MI in the black area involves two abnormal frequency jumping, where each is associated with the energy transfer between the spectral components that are not the nearest neighbours.
Namely,
\begin{eqnarray}
\tilde{\omega}\rightarrow k_1\tilde{\omega}\rightarrow k_2\tilde{\omega}, ~~~~k_1\geq3,~k_2\geq k_1+2.\label{eq-J3}
\end{eqnarray}
A specific case with $k_1=3$ and $k_2=5$ is shown in Fig. \ref{f10} with the parameters of the initial conditions represented by the yellow star in Fig. \ref{f12}.
Higher-order MI in the purple area shows a slight difference.
It involves, as the first step, energy transfer between the closest components ($\tilde{\omega}\rightarrow2\tilde{\omega}$). The subsequent step involves two abnormal frequency jumping which is similar with the progress described by Eq. (\ref{eq-J3}).
Namely,
\begin{eqnarray}
\tilde{\omega}\rightarrow 2\tilde{\omega}\rightarrow k_1\tilde{\omega}\rightarrow k_2\tilde{\omega}, ~~~~k_1\geq4,~k_2\geq k_1+2. \label{eq-J4}
\end{eqnarray}
The sharp edges of the cyan and black regions in Fig. \ref{f12} (b) are located on the red dashed line  $\beta=\beta_b=\pm\sqrt{6}a/2$.

\section{Conclusions}\label{Sec8}

In conclusion, we have studied the higher-order MI that involves abnormal frequency jumping during the nonlinear stage in the frame of multi-component NLSEs.
Based on the degeneracy analysis given by Eq. (\ref{eqzf}), we have studied systematically the case of three-component NLSEs which cannot reduce to the one- or two-component NLSEs.
In contrast to the two-component NLSEs, we have demonstrated in both focusing and defocusing regimes, the excitation and existence diagram of a class of nondegenerate ABs formed by nonlinear superposition between several fundamental breathers with the same unstable frequency but corresponding to different eigenvalues. Based on the eigenvalue analysis and the Hessian matrix analysis, we have presented the existence diagram of nondegenerate ABs on the ($\omega,\beta$) plane shown in Fig. \ref{f6}.
Starting with a pair of unstable frequency sidebands, we have shown in the higher-order MI, abnormal frequency jumping over the stable gaps between the instability bands in both defocusing and focusing regimes.
We have outlined the initial excitation diagram in Fig. \ref{f12}, which could be useful in experiments for observations of higher-order MI in multi-component Bose-Einstein condensates and multi-direction hydrodynamics.
Despite being highly complex, the numerical results have been confirmed by the exact solutions of multi-ABs of the multi-component NLSEs.

\section*{ACKNOWLEDGEMENTS}
The work is supported by the NSFC (Grants No. 12175178, and No. 12247103), and 2024 Northwest University excellent doctoral dissertation cultivation program (Grant No.YB2024007).

\begin{appendix}
\section{Eigenvalues of several cases of $N$-component NLSEs}\label{chi-expression}

(1) For $N=1$,  when $a_1=a$ and $\beta_1=\beta$, the expression for the eigenvalues can be obtained from  Eq. (\ref{eqchi}),
\begin{eqnarray}
\chi_1=\frac{1}{2}\left(-\omega-2\beta-\widehat{\chi_a}\right),
\chi_2=\frac{1}{2}\left(-\omega-2\beta+\widehat{\chi_a}\right),
\end{eqnarray}
where
\begin{equation}\label{chiin1}
\widehat{\chi_a}=\sqrt{\omega^2-4\sigma a^2}.
\end{equation}

(2) For $N=2$, we set $a_1=a_2=a$ and $\beta_1=-\beta_2=\beta$.
According to Eq. (\ref{eqchi}), we obtain the eigenvalues:
\begin{eqnarray}
\chi_1=\frac{1}{2}\left(-\omega-\widehat{\chi_a}\right),
\chi_2=\frac{1}{2}\left(-\omega+\widehat{\chi_a}\right), \\
\chi_3=\frac{1}{2}\left(-\omega-\widehat{\chi_b}\right),
\chi_4=\frac{1}{2}\left(-\omega+\widehat{\chi_b}\right),
\end{eqnarray}
where
\begin{equation}\label{chiin2}
\begin{split}
&\widehat{\chi_a}=\sqrt{\omega^2+4\beta^2-4\sigma a^2-4\mu}, \\
&\widehat{\chi_b}=\sqrt{\omega^2+4\beta^2-4\sigma a^2+4\mu},
\end{split}
\end{equation}
Here, parameters $\mu$ is
\begin{eqnarray}
&&\mu=\beta^2\omega^2-4\sigma a^2\beta^2+a^4,\nonumber
\end{eqnarray}

(3) For $N=3$, the parameters of the plane wave background are
\begin{equation}\nonumber
\begin{split}
a_1=a_2=a_3=a,~~ \\
\beta_1=-\beta_3=\beta,~\beta_2=0.
\end{split}
\end{equation}
Substituting the symmetric framework  into Eq. (\ref{eqchi}), we obtain the eigenvalues:
\begin{eqnarray}
\chi_1=\frac{1}{2}\left(-\omega-\widehat{\chi_a}\right),
\chi_2=\frac{1}{2}\left(-\omega+\widehat{\chi_a}\right), \\
\chi_3=\frac{1}{2}\left(-\omega-\widehat{\chi_b}\right),
\chi_4=\frac{1}{2}\left(-\omega+\widehat{\chi_b}\right), \\
\chi_5=\frac{1}{2}\left(-\omega-\widehat{\chi_c}\right),
\chi_6=\frac{1}{2}\left(-\omega+\widehat{\chi_c}\right),
\end{eqnarray}
where
\begin{equation}\label{chiin3}
\begin{split}
&\widehat{\chi_a}=\sqrt{\omega^2-\frac{4}{3}\left(3\sigma a^2-2\beta^2+\frac{\mu}{\nu}+\nu\right)}, \\
&\widehat{\chi_b}=\sqrt{\omega^2-\frac{4}{3}\left(3\sigma a^2-2\beta^2-\frac{1+i\sqrt{3}}{2}\frac{\mu}{\nu}-\frac{1-i\sqrt{3}}{2}\nu\right)}, \\
&\widehat{\chi_c}=\sqrt{\omega^2-\frac{4}{3}\left(3\sigma a^2-2\beta^2-\frac{1-i\sqrt{3}}{2}\frac{\mu}{\nu}-\frac{1+i\sqrt{3}}{2}\nu\right)}. \\
\end{split}
\end{equation}
Here, parameters $\mu$ and $\nu$ are
\begin{eqnarray}
&&\mu=\beta^4+(3\omega^2-12\sigma a^2)\beta^2+9 a^4,\nonumber\\
&&\nu=(\tau-\sqrt{\tau^2 - \mu^3})^{1/3},\nonumber
\end{eqnarray}
with
\begin{eqnarray}
\tau=\beta^6+(36\sigma a^2-9\omega^2)\beta^4-54 a^4\beta^2+27\sigma a^6.\nonumber
\end{eqnarray}

(4) For $N=4$, we set
\begin{equation}\nonumber
\begin{split}
a_1=a_2=a_3=a_4=a,~~ \\
\beta_1=-\beta_4=\beta,~\beta_2=\beta_3=0.
\end{split}
\end{equation}
Substituting into Eq. (\ref{eqchi}), we obtain the eigenvalues:
\begin{eqnarray}
\chi_1=\frac{1}{2}\left(-\omega-\widehat{\chi_a}\right),
\chi_2=\frac{1}{2}\left(-\omega+\widehat{\chi_a}\right), \\
\chi_3=\frac{1}{2}\left(-\omega-\widehat{\chi_b}\right),
\chi_4=\frac{1}{2}\left(-\omega+\widehat{\chi_b}\right), \\
\chi_5=\frac{1}{2}\left(-\omega-\widehat{\chi_c}\right),
\chi_6=\frac{1}{2}\left(-\omega+\widehat{\chi_c}\right),
\end{eqnarray}
where
\begin{equation}\label{chiin4}
\begin{split}
&\widehat{\chi_a}=\sqrt{\omega^2-\frac{4}{3}\left(4\sigma a^2-2\beta^2+\frac{\mu}{\nu}+\nu\right)}, \\
&\widehat{\chi_b}=\sqrt{\omega^2-\frac{4}{3}\left(4\sigma a^2-2\beta^2-\frac{1+i\sqrt{3}}{2}\frac{\mu}{\nu}-\frac{1-i\sqrt{3}}{2}\nu\right)}, \\
&\widehat{\chi_c}=\sqrt{\omega^2-\frac{4}{3}\left(4\sigma a^2-2\beta^2-\frac{1-i\sqrt{3}}{2}\frac{\mu}{\nu}-\frac{1+i\sqrt{3}}{2}\nu\right)}. \\
\end{split}
\end{equation}
Here, parameters $\mu$ and $\nu$ are
\begin{eqnarray}
&&\mu=\beta^4+(3\omega^2-10\sigma a^2)\beta^2+16 a^4,\nonumber\\
&&\nu=(\tau-\sqrt{\tau^2 - \mu^3})^{1/3},\nonumber
\end{eqnarray}
with
\begin{eqnarray}
\tau=\beta^6+(39\sigma a^2-9\omega^2)\beta^4-(60a^4+9\sigma a^2\omega^2)\beta^2+64\sigma a^6.\nonumber
\end{eqnarray}

Note that under the symmetric framework, the eigenvalue expressions for the 4- and 3-component NLSE systems are similar, differing only in the prefactors of the amplitude $a$. It can be seen from Eqs. (\ref{chiin3}) and (\ref{chiin4}).

\section{Hessian matrix analysis}\label{Hessian}

\begin{figure*}[htb]
\centering
\includegraphics[width=150mm]{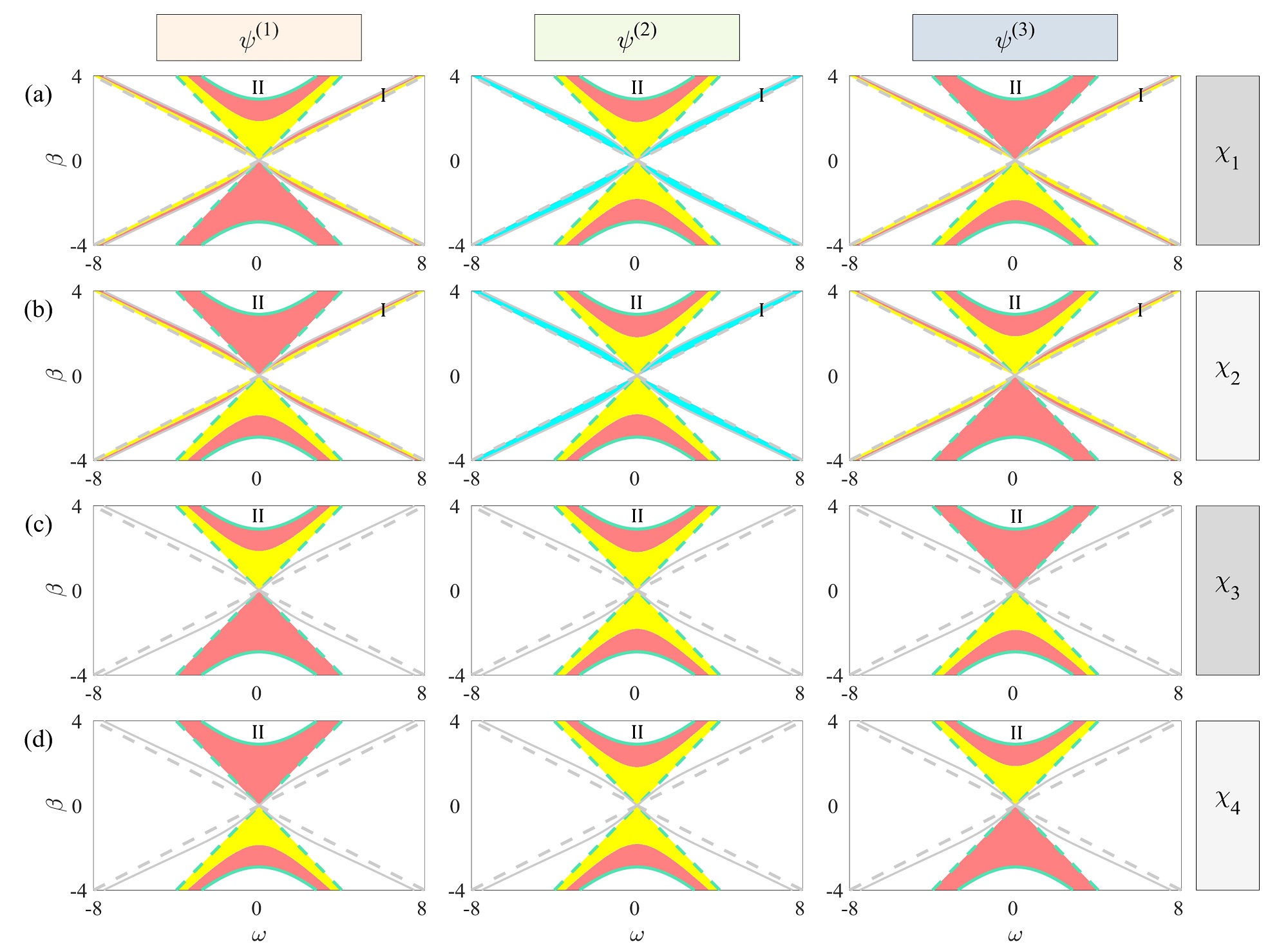}
\caption{ Existence areas of different types of ABs on the ($\omega,\beta$) plane in defocusing regimes ($\sigma=-1$)  for $\chi_{1,2,3,4}$, respectively.
The columns from left to right correspond to $\psi^{(1)}$, $\psi^{(2)}$ and $\psi^{(3)}$.
The pink, yellow and cyan areas correspond to dark, four-petal and bright ABs, respectively. The parameters are the same as in Figure \ref{f4}.
} \label{f13}
\end{figure*}

We obtain the diagram of AB structures by performing the Hessian matrix analysis \cite{VAB2021,Ling-2016}.
The Hessian matrix is defined by the  second-order partial derivatives at the central point $(\bm{\Gamma},\bm{\Omega})=(0,\pi)$,
\begin{eqnarray}
\mathcal{H}_j=\begin{pmatrix} |\psi^{(j)}/\psi_{0}^{(j)}|^2_{\bm{\Gamma}\bm{\Gamma}} & |\psi^{(j)}/\psi_{0}^{(j)}|^2_{\bm{\Gamma}\bm{\Omega}} \\ |\psi^{(j)}/\psi_{0}^{(j)}|^2_{\bm{\Gamma}\bm{\Omega}}  & |\psi^{(j)}/\psi_{0}^{(j)}|^2_{\bm{\Omega}\bm{\Omega}}\end{pmatrix}.\label{eqHM}
\end{eqnarray}
Three distinctive cases can be identified from Eq. (\ref{eqHM}): \\
(1) When $\det(\mathcal{H}_j)>0$, and $|\psi^{(j)}/\psi_{0}^{(j)}|^2_{\bm{\Gamma}\bm{\Gamma}}<0$, the Hessian is a negative definite matrix.
This implies that the special point is a maximum.
This corresponds to the `bright' structure.
The `bright' structure  shows that each cell has a high bump and small dips at each side of it.
\\
(2) When $\det(\mathcal{H}_j)<0$, the Hessian $\mathcal{H}_j$ is an indefinite matrix.
The centre of each cell in this case is a saddle point.
Each cell in the periodic pattern of vector components has a four-petal structure with two bumps and two dips symmetrically located around the centre.\\
(3) When $\det(\mathcal{H}_j)>0$, and $|\psi^{(j)}/\psi_{0}^{(j)}|^2_{\bm{\Gamma}\bm{\Gamma}}>0$, the Hessian is a positive definite matrix.
This referred to as the dark structure.
The dark structure exhibits that each cell has a dip surrounded by two small bumps on the sides.

\begin{figure*}[htb!]
\centering
\includegraphics[width=150mm]{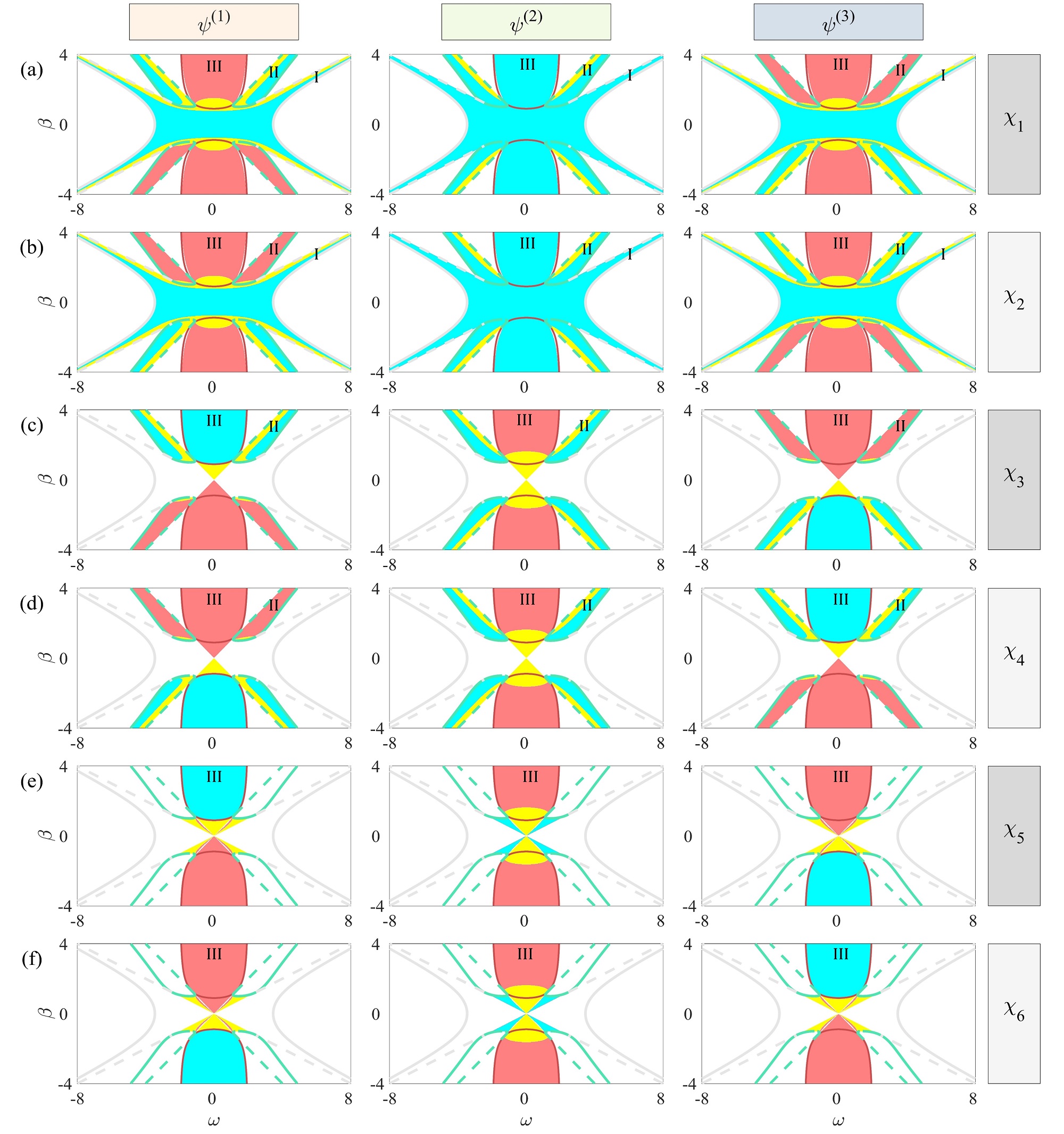}
\caption{Existence areas of different types of ABs on the ($\omega,\beta$) plane in focusing regimes ($\sigma=1$)  for $\chi_{1,2,3,4,5,6}$, respectively.
The columns from left to right correspond to $\psi^{(1)}$, $\psi^{(2)}$ and $\psi^{(3)}$.
The pink, yellow and cyan areas correspond to dark, four-petal and bright ABs, respectively. The parameters are the same as in Figure \ref{f5}.
} \label{f14}
\end{figure*}

Figures \ref{f13} and \ref{f14} depict the existence areas three qualitatively different types of ABs amplitude structures on the $(\omega,\beta)$ plane in the defocusing and focusing regime, respectively.
The pink, yellow and cyan areas correspond to dark, four-petal and bright ABs, respectively.

For the defocusing regime, when the amplitudes are equal, the bright AB structure is absent in the the defocusing two-component NLSE case \cite{VH_DMI}.
On the contrary, the three-component interaction can lead to the emergence of bright structures, as shown by the cyan areas in Fig. \ref{f13}.
For the focusing regime, when $\beta_1=\beta_2=\beta_3$, Eq. (\ref{eqsca}) shows that the three-component NLSE system reduces into a scalar one. This leads to that vector ABs show the same bright structure.
Thus, we omit the results corresponding to $\chi_{3,4,5,6}$.
The existence areas of the AB structures shown in Figs. \ref{f13} and \ref{f14} are completely consistent with the MI regions shown in Fig. \ref{f6}.

\section{multi-AB solutions}\label{multi-AB}
The general determinant form of the $M$-th order AB solutions via the  B\"{a}cklund transformation is:
\begin{eqnarray}
\psi^{(j)}_{M}=\psi_{0}^{(j)}{\det(\mathcal{G}^{(j)})}/{\det(\mathcal{G})},\label{vg1}
\end{eqnarray}
where
\begin{eqnarray}
\mathcal{G}^{(j)}&=&\begin{pmatrix} g^{(j)}_{1,1} & g^{(j)}_{1,2}& ... & g^{(j)}_{1,M}\\
g^{(j)}_{2,1} & g^{(j)}_{2,2}& ... & g^{(j)}_{2,M}\\
\vdots&\vdots& &\vdots&\\
g^{(j)}_{M,1} & g^{(j)}_{M,2}& ... & g^{(j)}_{M,M} \end{pmatrix},\\
\mathcal{G}&=&\begin{pmatrix} g_{1,1} & g_{1,2}& ... & g_{1,M}\\
g_{2,1} & g_{2,2}& ... & g_{2,M}\\
\vdots&\vdots& &\vdots&\\
g_{M,1} & g_{M,2}& ... & g_{M,M} \end{pmatrix}.
\end{eqnarray}
Here, $g^{(j)}_{m1,m2}$ and $g_{m1,m2}$ are the matrix elements of $\mathcal{G}^{(j)}$ and $\mathcal{G}$ in the $m1$-th row, $m2$-th column, respectively.
They are given by:
\begin{eqnarray}\label{ys}
\begin{split}
g_{m1,m2}&=\frac{\varphi_{m1}\varphi^*_{m2}}{\chi^*_{m2}-\chi_{m1}}+
\frac{\tilde{\varphi}_{m1}\tilde{\varphi}^*_{m2}}{\tilde{\chi}^*_{m2}-\tilde{\chi}_{m1}}
+\frac{\varphi_{m1}\tilde{\varphi}^*_{m2}}{\tilde{\chi}^*_{m2}-\chi_{m1}}\\
&+\frac{\tilde{\varphi}_{m1}+\varphi^*_{m2}}{\chi^*_{m2}-\tilde{\chi}_{m1}},\nonumber\\
g^{(j)}_{m1,m2}&=\frac{\chi^*_{m2}+\beta_j}{\chi_{m1}+\beta_j}
\frac{\varphi_{m1}\varphi^*_{m2}}{\chi^*_{m2}-\chi_{m1}}+
\frac{\tilde{\chi}^*_{m2}+\beta_j}{\tilde{\chi}_{m1}+\beta_j}
\frac{\tilde{\varphi}_{m1}\tilde{\varphi}^*_{m2}}{\tilde{\chi}^*_{m2}-\tilde{\chi}_{m1}}\\
&+\frac{\tilde{\chi}^*_{m2}+\beta_j}{\chi_{m1}+\beta_j}
\frac{\varphi_{m1}\tilde{\varphi}^*_{m2}}{\tilde{\chi}^*_{m2}-\chi_{m1}}+
\frac{\chi^*_{m2}+\beta_j}{\tilde{\chi}_{m1}+\beta_j}
\frac{\tilde{\varphi}_{m1}\varphi^*_{m2}}{\chi^*_{m2}-\tilde{\chi}_{m1}},
\end{split}
\end{eqnarray}
where $\ast$ denotes the complex conjugate, $\chi_{m}$ is the eigenvalue and $\tilde{\chi}_m=\chi_m+\omega_m$ $(m=1, 2, 3, ... M)$. Note that \begin{eqnarray}
\chi_{m1}=\chi_{m}|_{m=m1},~\tilde{\chi}_{m1}=\tilde{\chi}_{m}|_{m=m1},\nonumber\\
\chi_{m2}=\chi_{m}|_{m=m2},~\tilde{\chi}_{m2}=\tilde{\chi}_{m}|_{m=m2}.\nonumber
\end{eqnarray}

Similarly,
\begin{eqnarray}
\varphi(\chi_{m1})=\varphi(\chi_{m})|_{m=m1},~\varphi(\tilde{\chi}_{m1})=\varphi(\chi_{m})|_{m=m1},\nonumber\\
\varphi(\chi_{m2})=\varphi(\chi_{m})|_{m=m1},~
\varphi(\tilde{\chi}_{m2})=\varphi(\chi_{m})|_{m=m2}.\nonumber
\end{eqnarray}
where the functions $\varphi(\chi_m)$, and $\varphi(\tilde{\chi}_m)$  are given by
\begin{eqnarray}
\varphi(\chi_m)=\exp\{i\chi_m[(x-x_{m})+\frac{1}{2}\chi_m(t-t_{m})]\},\label{vg1}\\
\varphi(\tilde{\chi}_m)=\exp\{i\tilde{\chi}_m[(x-x_{m})+\frac{1}{2}\tilde{\chi}_m(t-t_{m})]\}.\label{vg1}
\end{eqnarray}
The real parameters $x_m$, and $t_m$ are the shifts in $x$ and $t$ of individual breathers, respectively.
For $M=1$, we obtain the fundamental vector AB solution of the Manakov equations. It is given, in simplified form, by Eq. (\ref{eqb}).

The $M$th-order solution corresponds to the nonlinear superposition of $M$ fundamental ABs, each associated with the parameters $(\chi_m, \omega_m, x_{m},t_{m})$, where $m=1,...,M$.
The space-time structure of a single AB in the superposition is directly determined by the parameters $\chi_{\textrm{set}}=\{\chi_{1}, ..., \chi_{M}\}$ and the frequencies $\omega_{\textrm{set}}=\{\omega_{1}, ..., \omega_{M}\}$.
The interaction between them (the spatiotemporal patterns of such multi-ABs) depends on the relative separations in both $x$ and $t$, i.e., $\Delta x_{\textrm{set}}=\{x_{1}, ..., x_{M}\}$ and $\Delta t_{\textrm{set}}=\{t_{1}, ..., t_{M}\}$.

The sixth-order solution $\psi^{(j)}_{M=6}$ involves six unstable frequencies $\omega_{\textrm{set}}=\{\omega_1,...,\omega_M\}=\{\tilde{\omega}, \tilde{\omega}, \tilde{\omega}, 3\tilde{\omega}, 3\tilde{\omega}, 5\tilde{\omega}\}$ is shown in Fig. \ref{f11}. The relative separations in both $x$ and $t$, i.e., $\Delta x_{\textrm{set}}=\{x_{1}, ..., x_{M}\}$ and $\Delta t_{\textrm{set}}=\{t_{1}, ..., t_{M}\}$ are given in the caption of Fig. \ref{f11}.

\end{appendix}

\end{document}